\documentclass[11pt]{article}%

\usepackage{colortbl,multirow,hhline}
\usepackage{graphicx}
\usepackage{multicol,multirow}
\usepackage{amsmath,amssymb,amsfonts, nccmath}
\usepackage{mathrsfs}
\usepackage{amsthm}
\usepackage{rotating}
\usepackage[title]{appendix}
\usepackage{float}
\usepackage[font=small,parskip=2pt]{caption}
\usepackage{titlesec}
\usepackage{gensymb}
\usepackage{float}
\usepackage{subfigure}
\usepackage{ifpdf}
\usepackage{mathtools}
\usepackage[T1]{fontenc}
\usepackage{newtxtext}
\usepackage{newtxmath}
\usepackage{textcomp}
\usepackage{tabularx}
\usepackage[dvipsnames]{xcolor}
\usepackage{lipsum}
\usepackage[super, square, comma, sort]{natbib}
\usepackage[resetlabels]{multibib}
\usepackage[top=2.5cm, bottom=2.5cm, left = 2.5cm, right = 2.5  cm]{geometry}
\usepackage[hidelinks]{hyperref}

\newcites{Supp}{References}

\usepackage{authblk}
\usepackage{titlesec}
 \usepackage[verbose]{placeins}
\definecolor{myblue}{cmyk}{0.7, 0.25, 0, 0.49}
\definecolor{myorange}{cmyk}{0, 0.8, 0.95, 0}
\definecolor{mygreen}{cmyk}{0.9, 0.3, 0.95, 0.3}

\theoremstyle{definition}

\numberwithin{equation}{section}
\title{Renewal equations for vector-borne diseases}
\author[1,2]{Cathal Mills*}
\author[1,2]{Tarek Alrefae}
\author[3]{William S. Hart}
\author[2,4]{Moritz U. G. Kraemer}
\author[5]{Kris V. Parag}
\author[3]{Robin N. Thompson}
\author[1,2]{Christl A. Donnelly}
\author[1]{Ben Lambert*}

\affil[1]{{Department of Statistics}, {University of Oxford}, {{Oxford}, {United Kingdom}}}
\affil[2]{{Pandemic Sciences Institute}, {University of Oxford},
{{Oxford}, {United Kingdom}}}
\affil[3]{{Mathematical Institute}, {University of Oxford}, {{Oxford}, {United Kingdom}}}
\affil[4]{{Department of Biology}, {University of Oxford}, {{Oxford}, {United Kingdom}}}
\affil[5]{{MRC Centre for Global Infectious Disease Analysis}, {Imperial College London}, {{London}, {United Kingdom}}}

\begin{document}
\maketitle
\small * Corresponding Author \\
\small E-mail: cathal.mills@linacre.ox.ac.uk (C.M) \\
\abstract{
During infectious disease outbreaks, estimates of time-varying pathogen transmissibility, such as the instantaneous reproduction number $R(t)$ or epidemic growth rate $r_t$, are used to inform decision-making by public health authorities. For directly transmitted infectious diseases, the renewal equation framework is a widely used method for measuring time-varying transmissibility. The framework uses information on the typical time elapsing between an infection and the offspring infections (quantified by the generation time distribution), and $R(t)$, to describe the rate at which currently infected individuals generate new infections. For diseases with transmission cycles involving hosts and vectors, however, renewal equation models have been far less used. This is likely due to difficulties in mechanistically defining  generation times that can capture the complexity of multi-stage, human-vector relationships. Here, using dengue as an example, we provide general renewal equations that are derived from first principles using age-structured systems of coupled partial differential equations across human and vector sub-populations. Our framework tracks the multi-stage transmission cycle over calendar time and across stage-specific ages, resulting in governing renewal equations that quantify how the rate at which new infections are generated from existing infections depends on stage-specific processes. The framework provides a foundation on which to base inferential frameworks for estimating $R(t)$ and $r_t$ for infectious diseases with multiple stages in the transmission cycle.

}
\section{Introduction}
In infectious disease epidemiology, renewal equations can be used to describe how new infections are generated from individuals infected in the past. The time period that elapses between a previous infection and a new infection caused by it is called the generation time, and the variation in this time period across individuals is captured by the generation time distribution \citep{svensson_note_2007}. The number of infections, on average, that a given infected individual would generate (should transmission conditions remain the same) throughout the course of their infection is known as the instantaneous reproduction number, $R(t)$ \citep{fraser_estimating_2007}. 

For directly transmitted (human-to-human) infectious diseases, renewal equation frameworks have found widespread use for estimating $R(t)$. The popularity is likely in part due to the relatively few assumptions needed to produce $R(t)$ estimates compared to other frameworks such as compartmental models. By fitting renewal equation models to time series of infection counts, or more commonly, case counts, it is possible to estimate $R(t)$, and these models form the basis for many popular approaches and software used to infer changes in transmission throughout outbreaks \citep{fraser_estimating_2007, cori_new_2013, thompson_improved_2019, abbott_estimating_2020, parag_improved_2021}. A renewal equation model for a directly transmitted infectious disease may be described by
\begin{align}\label{eq:ubiquituous_renewal}
i(t) &= R(t) \int_{0}^{\infty} i(t-\tau) w(\tau)d\tau,
\end{align}
where $i(t)$ is the number of new infections generated at time $t$, $R(t)$ is the instantaneous reproduction number at time $t$, and $w(\tau)$ is the probability density that describes the generation time distribution, resulting in $\int_0^\infty w(\tau) = 1$. Eq. \eqref{eq:ubiquituous_renewal} intuitively gives a process by which previous infections beget new infections with some time delay, and if $R(t)>1$, epidemic growth occurs. Consequently, the threshold of $R(t) > 1$ coincides with an instantaneous epidemic growth rate $r_t > 0$, where $r_t$ is the rate of change of the log infection incidence in calendar time. 

The renewal equation is most often implemented in discrete time to align with the temporal resolution of surveillance data \citep{ogi-gittins_simulation-based_2023}. Renewal models, as used for inference, also often include stochasticity (e.g. \citep{pakkanen_unifying_2023}). Misspecification of the generation time distribution can be a large potential source of error in this estimation of $R(t)$ \citep{parag_are_2022, park_importance_2022, parag_angular_2023, park_estimating_2024}, and given the popularity of $R(t)$ for informing public health policy and decision-making \citep{vegvari_commentary_2022}, the violation of the assumption of time-invariant generation times clearly has practical, real-world consequences.



For infectious diseases with a single population (equivalently, a single transmission stage) and z static generation time distribution, continuous renewal equations (such as eq. \eqref{eq:ubiquituous_renewal}) can be derived from a partial differential equation (PDE) representation of the infection-age-structured model of the infected population; this was first derived by \citet{kermack_contribution_1927} \citep{keyfitz_mckendrick_1997, champredon_equivalence_2018}. The renewal equation can also be used to estimate so-called state reproduction numbers, starting from an age-structured system with multiple host populations \citep{inaba_state-reproduction_2008}. 

\begin{figure}[H]
\centering
\includegraphics[scale = 0.685]{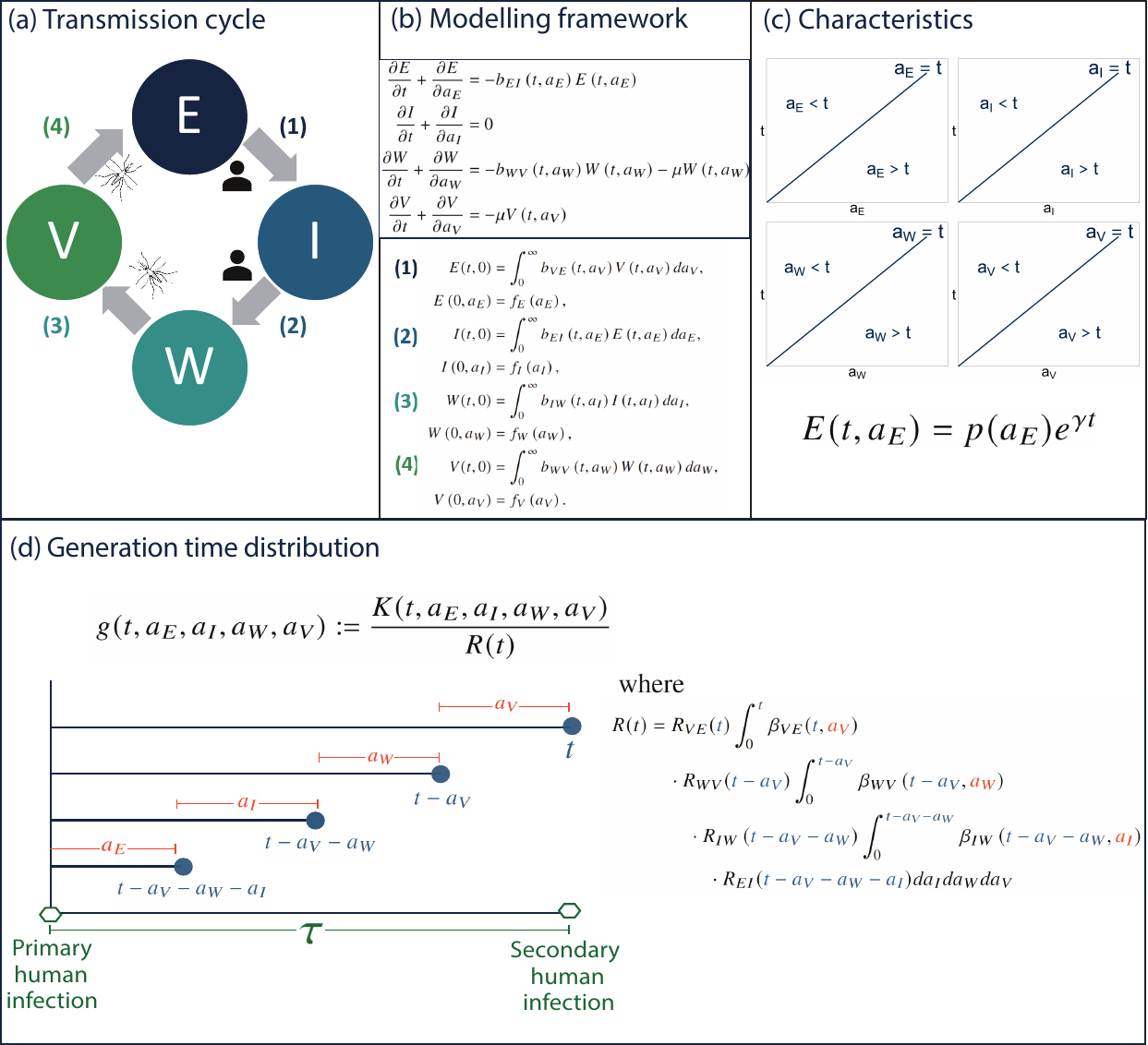}
\caption{\label{fig:figure_1} \footnotesize \textbf{Schematic of key concepts used to estimate $R(t)$ for vector-borne diseases}. a) The four-stage transmission cycle for dengue -- from exposed humans (E) to infectious humans (I) to exposed mosquitoes (W) to infectious mosquitoes (V). The boundary conditions of our modelling framework generate births into each sub-population. b) Our full modelling framework, eqs \eqref{eq:exposed_model_four} - \eqref{eq:initial_on_infectious_mosquitoes_model_four}, including the PDEs, and their boundary and initial conditions. c) Our methodology for solving the PDE system includes the method of characteristics and seeking a long-term solution of the form: $E(t, a_E) = p(a_E)e^{\gamma t}$ for the exposed humans. d) The generation time distribution at calendar time $t$, described by $g(t, a_E, a_I, a_W, a_V)$, is obtained by normalising the instantaneous kernel $K(t, a_E, a_I, a_W, a_V)$ using $R(t)$. $R(t)$ can be estimated using between-stage reproduction numbers and generation times using ages (orange) at intermediate calendar times (blue) when generation events (i.e. stage transitions) occur.}
\end{figure}

For vector-borne diseases, renewal equations are less commonly used as i) it is unclear whether the popular renewal equation can generalise to the multi-stage transmission cycle across various host and vector sub-populations (e.g. four stages for dengue across mosquito and human populations, Figure \ref{fig:figure_1}), and ii) the common assumption in renewal equations (as in eq. \eqref{eq:ubiquituous_renewal}) of a static generation time distribution is often unrealistic for climate-sensitive, vector-borne diseases. Under laboratory conditions, several mosquito-viral traits, such as the extrinsic incubation period, the rate of mosquito blood feeding, and mosquito density, have shown to be influenced by climate factors such as temperature \citep{chan_incubation_2012, mordecai_detecting_2017, mordecai_thermal_2019}. Also, the time taken for an infectious (or susceptible) mosquito to bite a susceptible (infectious) human will depend on susceptible human (mosquito) population levels. 

Existing approaches have used temperature data to parameterise temperature-dependent, time-varying generation times to estimate basic ($R_0$) and instantaneous reproduction numbers ($R(t)$). These approaches often involve a convolution of probability densities for the stages of the dengue transmission cycle, Ross-Macdonald assumptions for mosquito-borne pathogens, and/or parametric distributions for temperature-dependent generation times \citep{ross_prevention_1910, smith_ross_2012, siraj_temperature_2017, mordecai_thermal_2019, romeo-aznar_crowded_2024}. Such approaches, however, do not provide a mechanistically derived generation time distribution -- as the relative contributions of previously infected humans and mosquitoes to new human infections.

Here, we adopt a different approach by deriving renewal equations for a vector-borne disease starting from an age-structured system of coupled PDEs. To the best of our knowledge, this is the first derivation of renewal equations for a vector-borne disease. In doing so, we derive a mechanistically based generation time distribution which informs estimates of $R(t)$ from our renewal equation. Throughout, we use dengue as a case-study to motivate and develop our renewal equations, yet our approach can generalise to other mosquito-borne diseases (e.g. Zika, malaria, or yellow fever).
\section{Methods and results}
\subsection{Infection-age-structured human and mosquito populations} \label{sec:infection_structured_model_2}
\subsubsection*{Modelling framework} 
To illustrate calculation steps for our full model of the dengue transmission cycle (Section \ref{sec:exposed_and_infection_structured_model_4}), we start with a simplified abstraction of the dengue transmission cycle. We do not explicitly track the calendar times when a human (mosquito) becomes infectious after the viral infection event (i.e. after a mosquito bites a human). Our age-structured framework (further details of our framework are in Appendix \ref{sec:exposed_and_infection_structured_model_2_appendix}) describes both the infected mosquito and human populations using continuity equations; the McKendrick–von Foerster equation. \citep{mckendrick_applications_1925, murray_mathematical_2002}. 

Our system of age-structured populations is described by the following conservation equations:
\begin{align}
    \frac{\partial E}{\partial t}+\frac{\partial E}{\partial a_E} &=0,  \label{eq:exposed_model_2}\\
  \frac{\partial M}{\partial t}+\frac{\partial M}{\partial a_M} &=-\mu M(t,a_M),  \label{eq:mosquito_model_2}  
\end{align}
where $t$ denotes calendar time,  $E \geq 0$ and $M \geq 0$ denote the density of the infected human and mosquito populations respectively with corresponding ages $a_E \geq 0$ and $a_M \geq 0$, and $\mu>0$ is the (fixed) mortality rate in the infectious mosquito population. The variables $a_E$ and $a_M$ denote the time since entering the stage (not the age of a mosquito/human). For example, $E(t=3, a_E = 2)$ denotes the density of humans who were infected two time units (e.g. days) before calendar time $t = 3$. Each first-order linear equation, eqs.  \eqref{eq:exposed_model_2} -- \eqref{eq:mosquito_model_2}, posits that, because age $a_C$ (defined as time since entry to a sub-population $C$) and calendar time $t$ are measured in the same units, the rate at which a sub-population ages would equal the rate of population change with respect to time in the absence of any instantaneous death rate (i.e. the rate at which individuals leave a sub-population $C$). See \citet{murray_mathematical_2002} Chapter 1.7 for more details.

We close the system with the following boundary conditions:
\begin{align}
    E(t,0) &= \int_{0}^{\infty} b_{ME}(t,a_M) M(t,a_M) da_M,\label{eq:conservation_model_2}\\
    E(0,a_E) &= f_E(a_E),\\
    M(t,0) &= \int_{0}^{\infty} b_{EM}(t,a_E) E(t,a_E) da_E,\label{eq:conservation_model_2_m}\\
    M(0,a_M) &= f_M(a_M), \label{eq:initial_mosquito_model_2}
\end{align}
where $f_E(a_E)$ and $f_M(a_M)$ denote the initial densities of infected human and mosquito populations, and each $b_{CD}(t,a_C) \geq 0$ is a time- and age-dependent birth function that represents the rate at which new individuals in sub-population $D$ are generated at calendar time $t$ by an individual of age $a_C$ in sub-population $C$. We can solve this system using the method of characteristics -- a method that reduces our first-order linear PDE to a family of ODEs. 

\subsubsection*{Deriving a renewal equation and the time-varying reproduction number} 
To solve the system of PDEs, eqs \eqref{eq:exposed_model_2}-\eqref{eq:initial_mosquito_model_2}, we use i) the characteristic form of the infected mosquito population (e.g. Figure \ref{fig:figure_1}c), ii) the boundary condition on the mosquito population, eq. \eqref{eq:conservation_model_2_m}, and iii) and the characteristic form of the infected human population to derive the following renewal equation (see Appendix \ref{sec:exposed_and_infection_structured_model_2_appendix}):
\begin{equation}\label{eq:renewal_model_2}
     E(t,0) = \int_{0}^{t} \int_{0}^{t - a_M} b_{ME}(t,a_M)  e^{-\mu a_M}  b_{EM}(t-a_M,a_E) E(t-a_M-a_E,0) da_E da_M,
\end{equation}
where we have assumed that the epidemic is well-established, allowing us to focus on longer-term dynamics (such that $t \gg a_M$ and $t \gg a_E$). 

\begin{figure}[H]
\centering
\includegraphics[scale = 0.21]{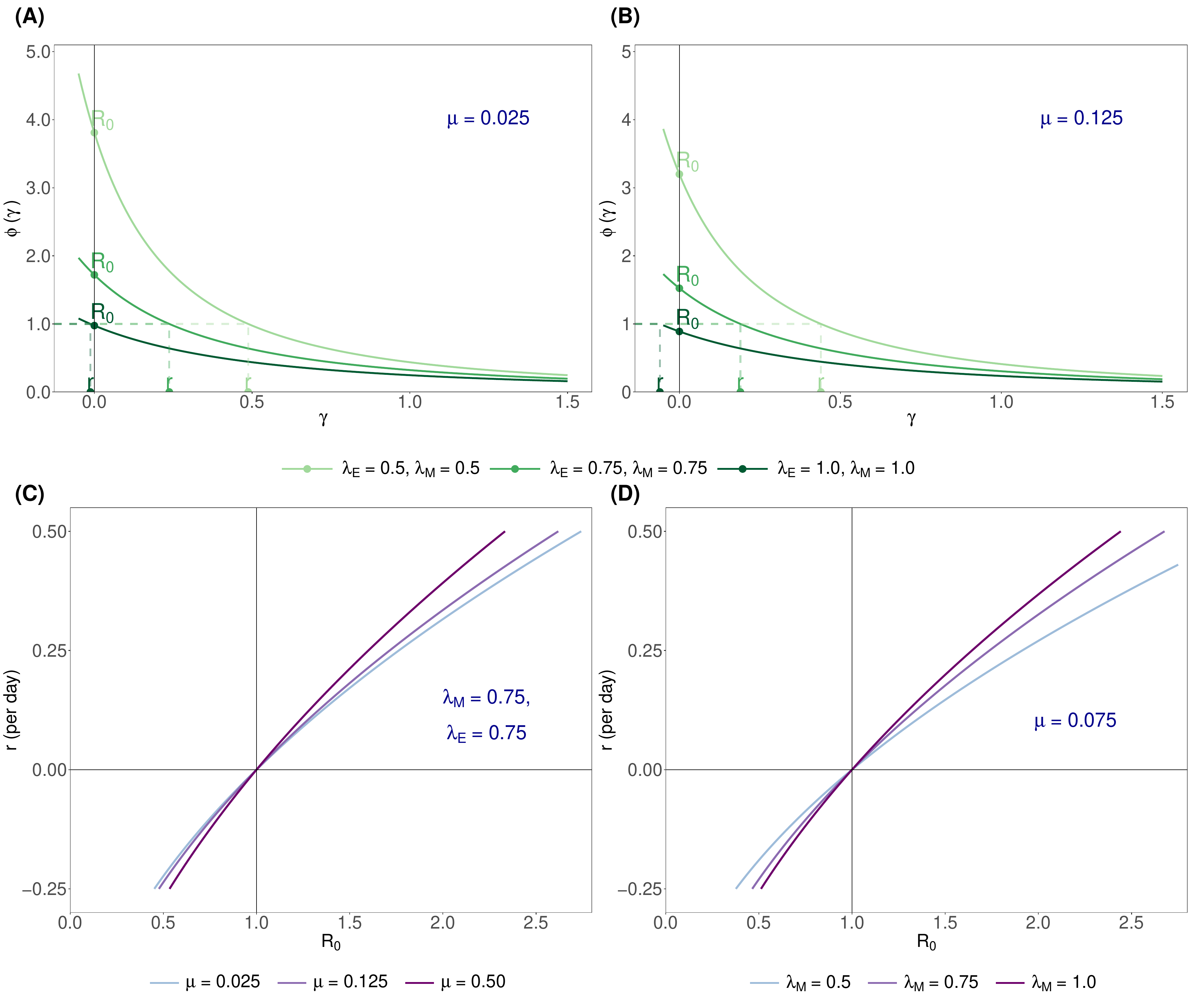}
\caption{\label{fig:R_r_analytical_plot} \footnotesize \textbf{$R_0$ and $r$ under different mosquito-human traits} for the modelling framework described by eqs. \eqref{eq:exposed_model_2}-\eqref{eq:conservation_model_2_m}. We assume $b_{EM}(t, a_E) = \exp({-\lambda_{E}a_E})$ and $b_{ME}(t, a_M) = \exp({-\lambda_{M}a_M})$. These assume that the birth processes decline only with time since infection. Panels \textbf{(A)} and \textbf{(B)}: We visualise how $R_0$ and $r$ are derived for varying infection rates ($\lambda_M$ and $\lambda_E$ for the production of infected humans and mosquitoes respectively) and mosquito death rates ($\mu$). $r = \{\gamma \text{ such that }\phi(\gamma) = 1\}$ and $R_0 = \phi(0)$. Panels \textbf{(C)} and \textbf{(D)}: The relationship between $r$ and $R_0$, eq. \eqref{eq:R_as_function_of_r} in Appendix \ref{sec:analytical_appendix}, is traced out for the modelling framework under three different mosquito death rates $\mu$ (left) and three fixed different mosquito infection rates $\lambda_M$ (right). These figures would be visualisations of the relationship between $R(t)$ and $r_t$, at each calendar time $t$, if we had not assumed time-independent birth processes $b_{EM}(t, a_E)$ and $b_{ME}(t, a_M)$.}
\end{figure}

Then, we seek a solution of the form: $E(t,a_E) = p(a_E) e^{\gamma t}$. This solution states that the age distribution of infected humans is altered by a factor which grows or decays with time depending on whether $\gamma > 0$ or $\gamma < 0$, respectively. Substituting the expression into eq. \eqref{eq:exposed_model_2}, we deduce that $p(a_E) = p(0) e^{-\gamma a_E}$. From eq. \eqref{eq:renewal_model_2}, we have
\begin{equation} \label{eq:phi_gamma_model_2}
    1 = \int_{0}^{t} \int_{0}^{t - a_M} b_{ME}(t,a_M)  e^{-\mu a_M}  b_{EM}(t-a_M,a_E) e^{-\gamma (a_E+a_M)}  da_E da_M := \phi(\gamma).
\end{equation}
This expression is defined for each calendar time $t$. From eq. \eqref{eq:phi_gamma_model_2}, we determine the instantaneous growth rate $r_t = \{\gamma \text{ such that }\phi(\gamma) = 1\}$. As $\phi(\gamma)$ is a monotonically decreasing function of $\gamma$, $\phi(0)$ is the critical threshold for epidemic growth. This is because $\gamma > 0$ when $\phi(0) > 1$ and $\gamma < 0$ when $\phi(0) < 1$ (See Figure \ref{fig:R_r_analytical_plot} A and B). 
\FloatBarrier
\vspace{1cm}
So, we define $R(t)$ as:
\begin{align}\label{eq:reproduction_number_model_2}
    R(t) &:= \phi(0) = \int_{0}^{t} \int_{0}^{t - a_M} b_{ME}(t,a_M)  e^{-\mu a_M}  b_{EM}(t-a_M,a_E) da_E da_M. 
\end{align}
$R(t)$ defines a time-varying human-to-human reproduction number because our renewal equation for the newly infected human population (eq. \eqref{eq:renewal_model_2}) has only a single birth state (for new human infections), and so the corresponding next-generation operator \citep{diekmann_mathematical_2000}, describing the number of new human infections produced by previously infected individuals, is:
\begin{align*}    
\mathbb{K} := \int_{0}^{t} \int_{0}^{t - a_M} b_{ME}(t,a_M)  e^{-\mu a_M}  b_{EM}(t-a_M,a_E) da_E da_M.
\end{align*}

The time-varying reproduction number $R(t)$ is the largest (and only) absolute eigenvalue of $\mathbb{K}$. So, eq. \eqref{eq:reproduction_number_model_2} defines the instantaneous human-to-human reproduction number $R(t)$ as, relying on only information up to time $t$, it is the average number of secondary human infections that an infected human would generate over their infected lifetime if the current conditions remained the same. Here, when discussing current conditions, these are the conditions that currently affect infection of humans -- conditions which are determined by the historical conditions experienced for the previous stages of the transmission cycle.
\subsubsection*{A closer look at $R_0$ and $r$} 
If the birth processes $b_{ME}(t,a_M)$ and $b_{EM}(t,a_E)$ do not change over calendar time, and we take the limit as $t \rightarrow \infty$, then eq. \eqref{eq:reproduction_number_model_2} defines a basic reproduction number $R_0$ with a corresponding intrinsic growth rate $r$. We can use simple exponential forms for birth processes, $b_{ME}(t,a_M) = \exp({-\lambda_{M}a_M})$ and $b_{EM}(t,a_E) = \exp({-\lambda_{E}a_E})$, to obtain analytical results. We illustrate these derivations in Figure \ref{fig:R_r_analytical_plot} for $R_0$ and $r$ (see also Appendix \ref{sec:analytical_appendix} for further details). The figure shows the familiar threshold relationship between $R_0$ and $r$ i.e. $R_0 > 1$ when $r > 0$. Also, faster declines in infection rates (from time since infection) and higher mosquito death rates both increase the exponential growth rate $r$ for a fixed $R_0 > 1$. Intuitively, this is because if, for example, mosquitoes have shorter lives, yet bequeath the same number of infections over their infected lifetime, then the epidemic must be growing faster. The figure also demonstrates that interventions which reduce $R_0$ by a fixed amount cause smaller changes in $r$ as $R_0$ increases.


\subsubsection*{Instantaneous kernels and generation time distributions} 
$R(t)$ is often written in terms of i) the instantaneous kernel or ii) the generation time distribution and the instantaneous growth rate. To obtain this form, we write eq. \eqref{eq:reproduction_number_model_2} as
\begin{equation}\label{eq:reproduction_number_model_2_2D}
\begin{aligned}
    R(t) &= \int_{0}^{t} \int_{0}^{t-a_M} b_{ME}(t,a_M)  e^{-\mu a_M} b_{EM}(t-a_M,a_E) da_E da_M,\\
    &=   \int_{0}^{t} \int_{0}^{t-a_M} K(t,a_M,a_E) da_E da_M,
\end{aligned}
\end{equation}
which suggests defining a 2-dimensional probability density, at calendar time $t$, for the generation time $g(t,a_M,a_E):=K(t,a_M,a_E)/R(t)$. This distribution controls the times between the primary and secondary human infections. As required for a generation time distribution, it is the normalised version of the instantaneous kernel $K(t, a_M, a_E)$ \citep{park_importance_2022, metz_dynamics_1986,wallinga_how_2007}. The kernel defines the average rate at which secondary human infections are generated at time $t$ by i) a mosquito infected $a_M$ time units ago and ii) a human that was of infection age $a_E$ at time $t-a_M$. So, $g(t,a_M,a_E)$ defines the relative contributions to secondary human infections at time $t$ of a mosquito of infection age $a_M$ at time $t$ and human of infection age $a_E$ at time $t-a_M$. This is similar to existing interpretations for directly transmitted infectious diseases, as the expected relative contribution to the force of infection by a human infected a specific number of time units ago \citep{breda_formulation_2012, park_importance_2022}. 

Using eqns. \eqref{eq:reproduction_number_model_2} and \eqref{eq:reproduction_number_model_2_2D}, we have:
\begin{equation}\label{eq:R_r_relationship}
\begin{aligned}
    \frac{1}{R(t)} &= \int_{0}^{t} \int_{0}^{t-a_M} g(t,a_M, a_E)  e^{-r_t( a_E + a_M)} da_E da_M,
\end{aligned}
\end{equation}
which shows how $g(t, a_M, a_E)$ controls the relationship between $r_t$ and $R_t$ (Figure \ref{fig:R_r_analytical_plot} C and D). 

Our renewal equation for the newly infected human population can now be written in terms of the generation time distribution and the instantaneous reproduction number:
\begin{align}
E(t,0) &= \int_{0}^{t} \int_{0}^{t-a_M} K(t,a_M,a_E) E(t-a_M-a_E, 0) da_E da_M \label{eq:renewal_2d_model_2_kernel}, \\
          &= R(t)\int_{0}^{t} \int_{0}^{t-a_M} g(t,a_M,a_E) E(t-a_M-a_E, 0) da_E da_M. \label{eq:renewal_2d_model_2_overall_gi}
\end{align}
The renewal equation features $E(t-a_M-a_E, 0)$, the humans previously infected at time $t-a_M-a_E$. The corresponding generation time distribution is described by $g(t,a_M,a_E)$ and is calculated using intermediate calendar times when generation events occur in the transmission cycle (see eq. \eqref{eq:model_2_time_varying_gi_meaning} below). We define a generation event to occur when a new individual is generated (or born) in a sub-population. 

The key to our generation time distribution is that it depends on i) the stages of the transmission cycle and ii) the calendar times at which they begin -- see Figure \ref{fig:figure_1}d for a visualisation.


\subsubsection*{Between-stage reproduction numbers} 
We can also define a reproduction number between stages, which gives the average number of infected humans generated per infected mosquito:
\begin{equation}\label{eq:model_2_between_reprod_gi_1}
\begin{aligned}
R_{ME}(t) &:=\int_{0}^{t} b_{ME}(t,a_M)  e^{-\mu a_M} d a_M, \\
\beta_{ME}(t,a_M) &:= \frac{b_{ME}(t,a_M) e^{-\mu a_M}}{R_{ME}(t)},
\end{aligned}
\end{equation}
where $1 = \int_{0}^{t} \beta_{ME}(t,a_M) d a_M$ and $\beta_{ME}(t,a_M)$ is the probability density, at calendar time $t$, of the generation time from infected mosquitoes to infected humans. $\beta_{ME}(t,a_M)$ is defined by normalising the instantaneous kernel between stages (i.e. the average rate at which human infections are generated at time $t$ by a mosquito infected $a_M$ time units ago). So, $\beta_{ME}(t,a_M)$ defines the relative contribution of mosquitoes of age $a_M$ to human infections at time $t$.

We also produce a reproduction number for the human-to-mosquito component of the dynamics: 
\begin{equation}\label{eq:model_2_between_reprod_gi_2}
\begin{aligned}
R_{EM}(t-a_M) &:=\int_{0}^{t-a_M} b_{EM}(t-a_M,a_E) da_E, \\
\beta_{EM}(t-a_M,a_E) &:= \frac{b_{EM}(t-a_M,a_E)}{R_{EM}(t-a_M)},
\end{aligned}
\end{equation}
where $1=\int_{0}^{t-a_M} \beta_{EM}(t-a_M,a_E) da_E$ and $\beta_{EM}(t-a_M,a_E)$ is the probability density, at calendar time $t-a_M$, of the generation time from infected humans to infected mosquitoes. So, $\beta_{EM}(t-a_M,a_E)$ describes the relative contributions of infected humans (to the generation of newly infected mosquitoes) that were of age $a_E$ at time $t-a_M$. 

Eqs \eqref{eq:model_2_between_reprod_gi_1}-\eqref{eq:model_2_between_reprod_gi_2} collectively result in a modification of eq. \eqref{eq:renewal_model_2} to

\begin{align}\label{eq:renewal_2d_model_2_betwee_compartment}
    E(t,0) &= R_{ME}(t) \int_{0}^{t} \beta_{ME}(t,a_M) R_{EM}(t-a_M) \int_{0}^{t - a_M} \beta_{EM}(t-a_M,a_E) E(t-a_M-a_E, 0) da_E da_M,
\end{align}
and also of eq. \eqref{eq:reproduction_number_model_2} to
\begin{equation}\label{eq:reproduction_number_model_2_geometric}
\begin{aligned}
    R(t) &= R_{ME}(t) \int_{0}^{t}  \beta_{ME}(t,a_M) R_{EM}(t-a_M) da_M\\
    &= R_{ME}(t) \mathbb{E}_{ME}[R_{EM}(t-a_M)],
\end{aligned}
\end{equation}
where $\mathbb{E}_{ME}[X(t-a_M)]:=\int_{0}^{t - a_M} X(t-a_M) \beta_{ME}(t,a_M) da_M$. This defines the reproduction number as expectations across the between-stage reproduction numbers. Eq. \eqref{eq:reproduction_number_model_2_geometric} is similar to a geometric average of the between-stage reproduction numbers for an entire system (e.g. \citep{keeling_modeling_2008, funk_comparative_2016, brouwer_why_2022}), and here we define a human-to-human reproduction number $R(t)$ which requires that a new human infection requires a two-stage process.

\subsubsection*{Relation to existing renewal equation frameworks} 
Our integral equation, eq. \eqref{eq:renewal_2d_model_2_overall_gi}, can be related to the ubiquitous renewal equation frameworks for directly transmitted infectious diseases (see Appendix \ref{appendix:model_2_relation}). 

Since $t - (a_M+a_E)$ occurs in our renewal equation, eq. \eqref{eq:renewal_2d_model_2_overall_gi}, we can define a new variable, $\tau = a_E + a_M$, measuring the overall delay between primary and secondary human infections. By a change of variables and an integration step, we produce a one-dimensional renewal equation:
\begin{equation}\label{eq:renewal_2d_model_2_additive_delay_inhomogeneous_gi}
     E(t,0) = R(t)\int_{0}^{t}  w(t,\tau) E(t-\tau,0)d\tau,
\end{equation}
where $w(t, \tau):= \int_{0}^{\tau} g(t, \tau - a_E, a_E) da_E$.

If we assume the generation time distribution is constant over calendar time, (i.e. $g(t,a_M,a_E) = g(a_M,a_E), \forall t$), we recover the familiar renewal equation \citep{fraser_estimating_2007, cori_new_2013, abbott_estimating_2020, parag_improved_2021}, as in eq. \eqref{eq:ubiquituous_renewal}: 
\begin{equation}\label{eq:renewal_2d_model_2_additive_delay_homogeneous_gi}
     E(t,0) = R(t)\int_{0}^{t}  w(\tau) E(t-\tau,0)d\tau.
\end{equation}
Assuming a constant generation time distribution over calendar time means that the ratio
\begin{align}
\label{eq:model_2_time_varying_gi_meaning}
    g(t,a_M,a_E) &= \frac{K(t,a_M,a_E)}{R(t)}  = \frac{b_{ME}(t,a_M)  e^{-\mu a_M} b_{EM}(t-a_M,a_E)}{\int_{0}^{t} \int_{0}^{t-a_M} b_{ME}(t,a_M)  e^{-\mu a_M}  b_{EM}(t-a_M,a_E) da_E da_M}, 
\end{align}
is fixed over calendar time ($t$). So, this typically implicit modelling assumption means that the relative expected contribution to secondary human infections of each pair of ages (i.e. times-in-states $(a_M, a_E)$) is constant over the course of an outbreak. This means $K(t, a_M, a_E)$ scales linearly over calendar time $t$ with the expected overall number of new infections generated (across all age combinations of sub-populations) per initially infected human, $R(t)$.

The assumption may be unrealistic for dengue or any vector-borne disease with a multi-stage transmission cycle. For example, higher temperatures may reduce the extrinsic incubation period of dengue \citep{chan_incubation_2012}, and/or changes in the host and vector population sizes (which may themselves have temperature and precipitation dependencies) could alter the distribution of times between generation events. 

Eq. \eqref{eq:renewal_2d_model_2_additive_delay_homogeneous_gi} formalises the relationship between our framework and existing renewal equations (used in e.g. \citep{fraser_estimating_2007, cori_new_2013, abbott_estimating_2020, parag_improved_2021}). The existing renewal equations are a special case of our general framework with an additional assumption of a time-invariant generation time distribution. 
\subsection{Exposure-and-infection-age-structured populations of human and mosquitoes}\label{sec:exposed_and_infection_structured_model_4}
\subsubsection*{Modelling framework} 
We now extend the above framework to a transmission cycle with four sub-populations which includes exposed and infectious sub-populations in the host and vector populations. This structure accounts for the intrinsic and extrinsic latent periods (i.e. the time taken from mosquito bite to the host or mosquito respectively becoming infectious). The system, consisting of densities of an exposed ($E$) and infectious ($I$) human population, and an exposed ($W$) and infectious ($V$) mosquito population, is subject to the following conservation equations (e.g. Chapter 1.7 in \citet{murray_mathematical_2002}):
\begin{align}
\frac{\partial E}{\partial t}+\frac{\partial E}{\partial a_E} &= -b_{E I}\left(t, a_{E}\right) E\left(t, a_{E}\right), \label{eq:exposed_model_four}\\
\frac{\partial I}{\partial t}+\frac{\partial I}{\partial a_I} &= 0, 
\label{eq:infectious_model_four}
\\
\frac{\partial W}{\partial t}+\frac{\partial W}{\partial a_W} &= -b_{W V}\left(t, a_{W}\right) W\left(t, a_{W}\right) -\mu W\left(t, a_{W}\right),
\label{eq:mosquito_exposed_model_four} 
\\
\frac{\partial V}{\partial t}+\frac{\partial V}{\partial a_V} &=-\mu V\left(t, a_{V}\right), \label{eq:mosquito_infectious_model_four}
\end{align}
where again each $b_{CD}(t,a_C) \geq 0$ is a time- and age-dependent birth function that represents the rate at which new individuals in sub-population $D$ are generated at calendar time $t$ by an individual of age $a_C$ in sub-population $C$. $\mu \geq 0$ is the mosquito death rate for both exposed and infectious mosquitoes, which we assume is age- and infection-status-independent \citep{lambert_meta-analysis_2022}.

We close the system with the following initial and boundary conditions
\begin{align}
E(t, 0) &=\int_0^\infty b_{VE}\left(t, a_V\right) V\left(t, a_V\right) d a_V, \label{eq:boundary_on_exposed_model_four}\\
E\left(0, a_E\right) &= f_E\left(a_E\right), \\
I(t, 0) &=\int_0^{\infty} b_{E I}\left(t, a_E\right) E\left(t, a_E\right) d a_E, \label{eq:boundary_on_infectious_model_four} \\
I\left(0, a_I\right) &= f_I\left(a_I\right), \\
W(t, 0) &=\int_0^{\infty} b_{IW}\left(t, a_I\right) I\left(t, a_I\right) d a_I, \label{eq:boundary_on_exposed_mosquitoes_model_four} \\
W\left(0, a_W\right) &=f_W\left(a_W\right),  \\
V(t, 0) &=\int_0^{\infty} b_{WV}\left(t, a_W\right) W\left(t, a_W\right) d a_W, \label{eq:boundary_on_infectious_mosquitoes_model_four} \\
V\left(0, a_V\right) &=f_V\left(a_V\right).  \label{eq:initial_on_infectious_mosquitoes_model_four}
\end{align}
where $f_E(a_E), f_I(a_I), f_W(a_W),$ and $f_V(a_V)$ are the initial densities for the exposed human, infectious human, exposed mosquito, and infectious mosquito populations respectively. An intermediate system with three sub-populations is provided in Appendix \ref{sec:exposed_and_infection_structured_model_3} and additional derivation details for this four-compartment framework are provided in Appendix \ref{sec:exposed_and_infection_structured_model_4_appendix}.


\subsubsection*{Deriving a renewal equation and the time-varying reproduction number} 
We can derive the following renewal equation for newly infected humans:
\begin{equation}\label{eq:model_four_renewal_equation_on_infectious_in_manuscript}
\begin{aligned}
E(t, 0) &=\int_0^t \int_0^{t-a_V}  \int_0^{t-a_V-a_W}  \int_0^{t-a_V - a_W - a_I} b_{VE}\left(t, a_V\right)   b_{WV}\left(t-a_V, a_W\right) b_{IW}\left(t-a_V - a_W, a_I\right)  \\ & \quad \quad \cdot b_{E I}\left(t-a_V - a_W - a_I, a_E\right) e^{-\mu a_V} e^{-\int_0^{a_W} b_{WV}(t-a_V, r) d r - \mu a_W}    \\ & \quad \quad \quad \cdot E(t-a_V - a_W - a_I - a_E,0) e^{-\int_{0}^{a_E} b_{EI}(t-a_V - a_W - a_I,s) ds} d a_E  d a_I  d a_W   d a_V, & 
\end{aligned}
\end{equation}
\normalsize
where we have again assumed long-term epidemic dynamics (such that $t \gg a_V$, $t \gg a_W$, and $t \gg a_I$, $t \gg a_E$). We again try a separable solution (to derive analytical results) of the form $E(t,a_E) = p(a_E) e^{\gamma t}$. To have a separable form for eq. \eqref{eq:model_four_renewal_equation_on_infectious_in_manuscript}, we must assume that $b_{E I}\left(t, a_{E}\right) = b_{E I}\left(a_{E}\right) \forall t$. This amounts to assuming that the rate at which an exposed individual becomes infectious is independent of calendar time ($t$) and depends only on time since exposure ($a_E$). Then, we can derive a formula for the time-varying human-to-human reproduction number:
\begin{equation}
\begin{aligned}
\label{eq:reprod_number_model_four_in_terms_of_density_in_manuscript}
R(t) &= \int_0^t \int_0^{t-a_V}  \int_0^{t-a_V-a_W}  \int_0^{t-a_V - a_W - a_I} b_{VE}\left(t, a_V\right)   b_{WV}\left(t-a_V, a_W\right) b_{IW}\left(t-a_V - a_W, a_I\right)  \\ & \quad \quad\cdot b_{E I}\left(a_E\right) e^{-\mu a_V} e^{-\int_0^{a_W} b_{WV}(t-a_V, r) d r - \mu a_W} e^{-\int_{0}^{a_E} b_{EI}(s) ds} \\ & \quad \quad \quad d a_E  d a_I  d a_W   d a_V \\ \\
&= \int_0^t \int_0^{t-a_V}  \int_0^{t-a_V-a_W}  \int_0^{t-a_V - a_W - a_I}K(t,a_E, a_I,a_W,a_V) d a_E  d a_I  d a_W   d a_V,
\end{aligned}
\end{equation}
\normalsize
which suggests the following definition of a four-dimensional probability density
\begin{equation}\label{eq:gi_model_4}
g(t,a_E, a_I,a_W,a_V):=\frac{K(t,a_E, a_I,a_W,a_V)}{R(t)},
\end{equation}
for the generation time at calendar time $t$, derived by normalising the instantaneous kernel.

Again, the time-varying reproduction number $R(t)$ is an instantaneous reproduction number for the infected human population, and defines the expected number of secondary human infections produced at calendar time $t$ by an infected human, if the conditions that affect human infections at time $t$ were to remain the same. The instantaneous kernel $K(t,a_E, a_I,a_W,a_V)$ is the expected rate at which secondary human infections are generated at calendar time $t$ across individuals of stage-specific ages $(a_E, a_I,a_W,a_V)$; these ages are the times-since-state-entry at the calendar times when each sub-population plays their role in the transmission cycle. 

At each calendar time $t$, the generation time distribution describes the relative contributions to the current number of new human infections generated by previously exposed humans, previously infectious humans, previously exposed mosquitoes, and currently infectious mosquitoes of specific ages (i.e. the times spent in compartments until each population plays their role in generation events -- see Figure \ref{fig:figure_1}d). 

We can then rewrite eq. \eqref{eq:model_four_renewal_equation_on_infectious_in_manuscript} to produce a more familiar renewal equation:
\begin{equation}
\begin{aligned} 
\label{eq:renewal_4d_model_four_gi}
E(t, 0) &= 
 R(t) \int_0^t \int_0^{t-a_V}  \int_0^{t-a_V-a_W}  \int_0^{t-a_V - a_W - a_I}  g(t,a_E, a_I,a_W,a_V) \\ & \quad \quad \cdot E(t-a_V - a_W - a_I - a_E,0) d a_E  d a_I  d a_W   d a_V .  
 \end{aligned}
 \end{equation}

\subsubsection*{Between-stage reproduction numbers} 
As before, we can define reproduction numbers and generation time distributions between stages (see Appendix \ref{appendix:between_comp}). This allows us to rewrite eq. \eqref{eq:reprod_number_model_four_in_terms_of_density_in_manuscript} as an expectation of reproduction numbers across stages of the transmission cycle:
\begin{align}
R(t) &=R_{V E}(t)\mathbb{E}_{V E}\left[R_{WV}(t-a_V) \mathbb{E}_{W V}\left[ R_{IW}(t-a_V-a_W) \mathbb{E}_{I W} \left[R_{E I}\left(t-a_V-a_W-a_I\right)  \right] \right]\right], 
\end{align}
where each $\mathbb{E}_{AB}[R_{CD}(t - a_A)] :=\int_{0}^t R_{CD}(t-a_A) \beta_{AB}(t,a_A) da_A$ and each $R_{CD}(t)$ defines the average number of individuals in sub-population $D$ produced by an individual in sub-population $C$ if conditions were to remain the same as at time $t$. The between-stage reproduction numbers are evaluated with upper limits determined by the calendar times at which the subsequent stage of the transmission cycle begins. The expectations are taken with respect to the generation time distribution ($\beta_{AB}$) between stages of the next stage of the transmission cycle. The generation time distributions between stages are evaluated using ages of the sub-populations at the calendar times at which onward generation events in the transmission cycle occur (see Figure \ref{fig:figure_1}d: $a_V$ at time $t$, $a_W$ at time $t-a_V$, and $a_I$ at time $t-a_V-a_W$).

\begin{figure}[H]
    \centering
    \includegraphics[scale=0.1475]{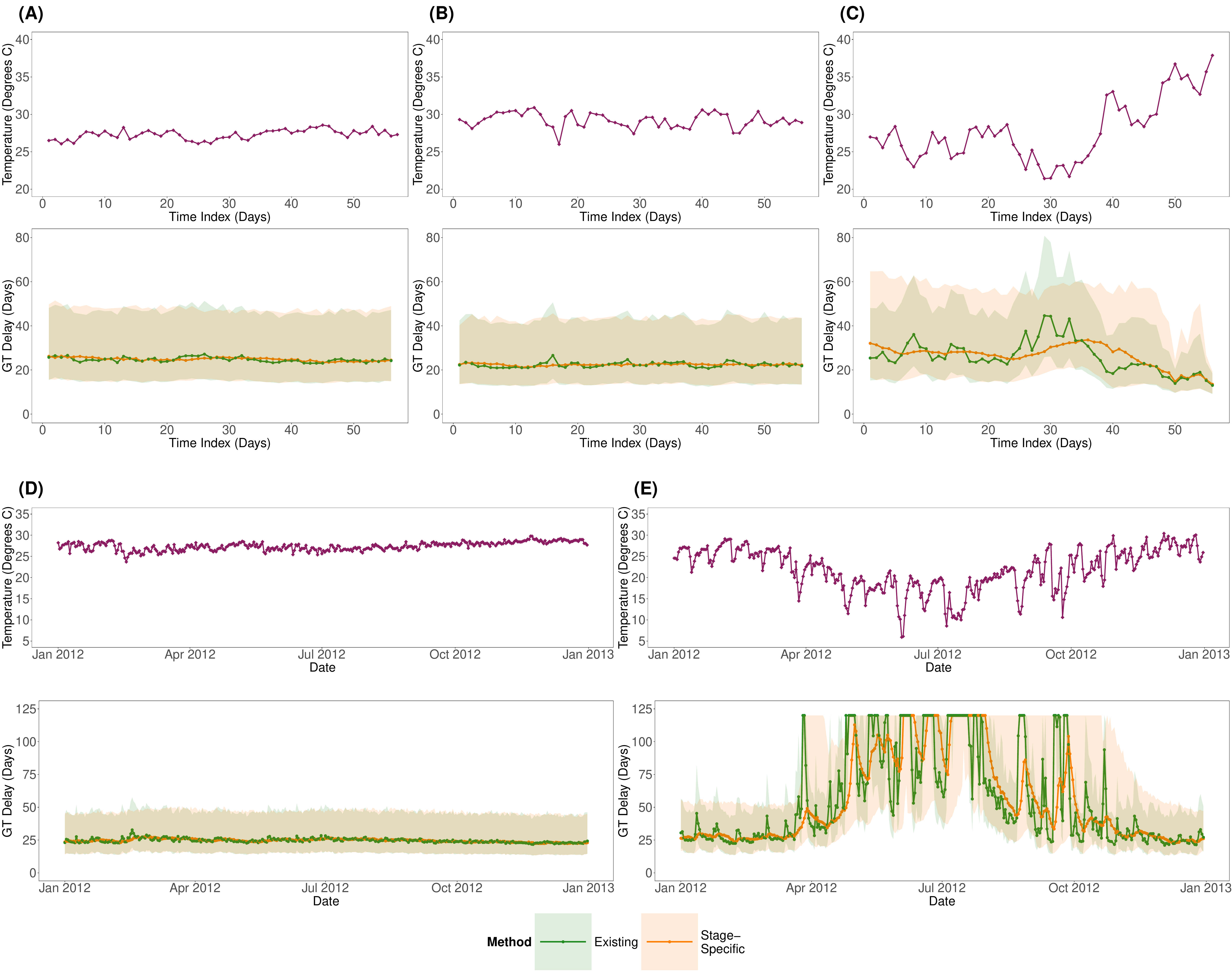}
    \caption{\textbf{Generation time (GT) distributions under temperature-dependent conditions:} For the four-stage transmission cycle of dengue, we illustrate GT distributions at different calendar times. In these toy examples, the distributions were estimated using Monte Carlo sampling (see Appendix \ref{appendix:application}). We assumed temperature-dependent GT distributions and used daily temperature data and either a stage-specific (orange) or the existing (green) approach. The filled circles indicate the medians of the 2000 Monte Carlo samples, and the shaded regions indicate the central 95\% intervals of the samples. The stage-specific approach tracks the calendar times at which the generation events of the transmission cycle occur. The existing approach allows the GT distribution to shift at each calendar time $t$ based on the temperature observed on that day. For each approach, we assumed the same temperature dependencies and probability distributions (for the transmission cycle stages) that were outlined by \citet{siraj_temperature_2017} (see Appendix \ref{appendix:application}). GT distributions were estimated using daily mean temperature (purple) data for (A) Ang Mo Kio, Singapore (December 2023 to June 2024), B) São João do Tauape, Brazil (November 2011 to May 2012), (C) a simulated environment (using a Gaussian random walk with noise $\epsilon_t \sim N(0, 4)$, and maximum and minima of 20 \degree C and 45 \degree C respectively), (D) São João do Tauape, Brazil (January 2012 to December 2012), and (E) Foz do Iguaçu, Brazil (January 2012 to December 2012)\citealp{hersbach_era5_2020,meteorological_service_singapore_historical_2024}. The upper plots, (A)-(C), show GT distributions over 56 days (8 weeks), while the lower plots, (D)-(E), show GT distributions over one year. We used different y-axes for the plots in panels (A)-(C) vs (D)-(E) for legibility. For panel (E), we bounded the maximum GT delay (y-axis) at 120 days, consistent with the approach taken by \citet{siraj_temperature_2017}.
\label{fig:application_gis_plot}}
\end{figure}
\subsubsection*{Application to simulated and real-world data}

We now investigate the importance of tracking the calendar times of the transmission cycle when estimating the generation time distribution. For simplicity, this example uses the assumptions governing how stage durations depend on temperature outlined by \citet{siraj_temperature_2017} (see Appendix \ref{appendix:application}). 

We used daily mean temperature data from Ang Mo Kio (a town in Singapore), Foz do Iguaçu (a city in Brazil), and São João do Tauape (a neighbourhood in the city of Fortaleza, Brazil), as well as simulated daily temperature data \citealp{meteorological_service_singapore_historical_2024, hersbach_era5_2020}. Locations were selected due to known circulation of dengue viruses. We investigated differences in generation time distributions when using a stage-specific approach versus one which ignores the calendar times of the stages of the transmission cycle. In each approach, we estimated the generation time distribution assuming the four-stage transmission cycle (Figure \ref{fig:figure_1}a) and Monte Carlo sampling (see Appendix \ref{appendix:application}). Our Monte Carlo approach can be described by forming $\tau = a_E + a_I + a_W + a_V$ as the delay between primary and secondary human infections, and then by sampling $a_E$, $a_I$, $a_W$, and $a_V$ from their assumed between-stage distributions; repeating this many times produces an estimate of $w(t, \tau$). 

For i) the comparator approach which ignores the calendar times at which stages of the transmission cycle occur, the generation time distribution is changed every day based on the temperature observed that day. We sampled $a_E^{i} \sim \beta_{EI}(t, a_E)$, $a_I^{i} \sim \beta_{IW}(t, a_W)$, $a_W^{i} \sim \beta_{WV}(t, a_W)$, and $a_V^{i} \sim \beta_{VE}(t, a_V)$ based on the temperature observed at the same calendar time $t$. For ii) our stage-specific approach (a discrete version of eq. \eqref{eq:gi_model_4}), the generation time distribution was based on temperatures observed for calendar times when the estimated generation events occurred. So, using the sampling distributions (between stages) outlined by \citet{siraj_temperature_2017}, we sampled $a_V^{i} \sim \beta_{VE}(t, a_V)$, $a_W^{i} \sim \beta_{WV}(t - a_V^{i}, a_W)$, $a_I^{i} \sim \beta_{IW}(t - a_V^{i} - a_W^{i}, a_W)$, and $a_E^{i} \sim \beta_{EI}(t-a_V^{i}-a_W^{i}-a_I^{i}, a_E)$. Then, for either approach, $\tau^{i} = a_E^{i} + a_I^{i} + a_W^{i} + a_V^{i}$ 

We investigated short time periods (eight weeks, Figure \ref{fig:application_gis_plot} A-C) where dengue was likely to circulate and longer time periods (an entire year, Figure \ref{fig:application_gis_plot} D-E) where dengue was less likely to circulate (due to winter temperatures). To illustrate the effects of a volatile environment, our simulated daily temperature data was based on a Gaussian random walk (with noise $\epsilon_t \sim N(0, 4)$, and reflecting boundaries at 20 \degree C and 45 \degree C).

For the eight-week period, in the applications to real-world temperature data (Figure \ref{fig:application_gis_plot} A-B, Figures \ref{fig:amk_both_gi_histogram} – \ref{fig:sjt_both_gi_density}), there were similar trends in the estimated generation time distributions across the two methods. However, the higher variability in the simulated temperature data (Figure \ref{fig:application_gis_plot}C,  Figures \ref{fig:sim2_both_gi_histogram} – \ref{fig:sim2_both_gi_density}) resulted in notable differences in the evolution of the generation time distributions. For the year-long period, in São João do Tauape (Figure \ref{fig:application_gis_plot} D), the generation time distribution from both methods did not fluctuate wildly over calendar time since the temperature was relatively stable temperature conditions. In contrast, the volatile temperatures observed in the city of Foz do Iguaçu (Figure \ref{fig:application_gis_plot} E) resulted in unrealistically long generation times. These results hint that at lower temperatures, transmission could be slowed by temperature-mediated effects. However, these estimates are the result of extrapolating laboratory results directly into the field, which likely overstates the impact of temperature on transmission (see Discussion). Changes in temperature are also likely correlated with rainfall amounts, resulting in changes in the mosquito population dynamics which may dominate the temperature effects. 

Across the five settings (Figure \ref{fig:application_gis_plot} A-E), the stage-specific approach acted like a smoothed average versus the existing approach as the generation time distribution was based on temperatures observed throughout the transmission cycle. These differences are most pronounced in the simulated setting (Figure \ref{fig:application_gis_plot} C) and Foz do Iguaçu (Figure \ref{fig:application_gis_plot} E) when temperatures varied more widely.
\section{Discussion}
Transmission of mosquito-borne pathogens is sensitive to environmental fluctuations, particularly to changes in temperature, and elements of their transmission cycle respond differently to these variations \citealp{chan_incubation_2012, mordecai_detecting_2017, huber_seasonal_2018, monteiro_dengue_2019}. This temporal variation means that there is no fixed distribution governing the time it takes one human infection to cause (through the vector) a subsequent human infection. This creates issues for traditional renewal frameworks which assume that this so-called generation time distribution is fixed. Here, we derive from first principles a renewal equation that applies to mosquito-borne diseases, which naturally accounts for temporal variation in each element of the transmission cycle. Time-varying reproduction numbers and epidemic growth rates emerge from our renewal equation framework, and this provides a theoretically grounded basis for estimating these important metrics of pathogen transmission rates.

Our analysis shows that the generation time distribution is likely to be transient, changing during a mosquito-borne disease outbreak in response to varying environmental conditions. The generation time distribution corresponding to secondary infections arising at any given calendar time depends on both current and historical environmental conditions. This dependence on history is because for transmission to occur, a full sequence of stages of the transmission cycle must be completed, and each of these stages could be exposed to different conditions. Our generation time distribution at a given time then represents the uncertainty in the typical time between parent and daughter infections if conditions were as they were historically. The mechanistic, history-dependent nature of our generation time distribution contrasts with existing less mechanistic approaches which more heuristically allow the generation time distribution to change in response to only current conditions \citealp{siraj_temperature_2017}. Our empirical results hint that our approach predicts less temporal variation in the generation time distribution than approaches based only on current conditions because the dependence on historical conditions likely acts to smooth its variation over time. Our generation time distribution also differs to other less mechanistic approaches which do not directly normalise the relative contributions of different sub-population ages (i.e. the instantaneous kernel), and instead only respond to observed temperature values -- ignoring other sources of temporal variation for elements of the transmission cycle \citealp{codeco_estimating_2018}

Ours and other approaches to modelling mosquito-borne pathogen transmission depend on assumptions about how elements of the pathogen and mosquito lifecycle respond to environmental conditions, and these are largely drawn from laboratory studies. Many such studies have shown that variations in temperature substantially affect the time spent in particular lifecycle stages. For malaria, Zika, dengue, and a range of arboviruses, the time taken for an infected mosquito to become infectious is hastened by higher environmental temperatures in the laboratory. Mosquitoes also die more rapidly at elevated laboratory temperatures \citealp{yang_assessing_2009, chan_incubation_2012,mordecai_thermal_2019}. These laboratory studies underpin a wide range of modelling studies which extrapolate how mosquito-borne pathogen transmission may respond to changes in temperature (e.g. \citealp{siraj_temperature_2017, chen_indian_2024}). However, more work is required to understand whether these results from laboratory studies hold in the wild: mosquitoes, like all insects, are cold-blooded and rely on their environment to thermoregulate meaning they adjust their behaviour in response to changes in environmental conditions. Since the laboratory mosquitoes are limited in their environment, they are likely limited in their ability to respond to temperature changes. Laboratory mosquitoes are often highly inbred which may also make them more fragile to temperature changes. This means that laboratory studies likely overstate the impacts of temperature on disease transmission, and field or semi-field studies are crucial. 

Despite recent advances in understanding temperature-dependent transmission \citealp{shocket_transmission_2020, tesla_temperature_2018, mordecai_detecting_2017}, we still do not have a comprehensive understanding of when, where, and why results from laboratory studies may transfer to the field. This is essential as there are many environmental and biological factors other than temperature that may affect the transmission of pathogens. For example, the rate at which mosquitoes become infected and infect humans depends on the population size of mosquitoes (and susceptible humans). Mosquito population dynamics are driven by the availability of water sources due to the dependence of the initial mosquito life stages on these. So, rainfall also affects the rate of pathogen transmission, although this aspect is typically ignored in existing renewal-type models of mosquito-borne infection transmission. Our model considers pathogen dynamics and only the dynamics of infected mosquitoes, and we could straightforwardly link it to the overall mosquito population dynamics by allowing the birth rate of infections in either humans or mosquitoes to depend on these. Modelling how mosquito population dynamics depend on both temperature and rainfall would then allow forward projections of pathogen transmission rates to be made.

Our framework is based on a deterministic model of pathogen dynamics, which models the density of each life stage by continuous variables. Because of this, it is likely most appropriate for modelling the dynamics of larger populations. Particularly for small populations, it would be useful to develop a stochastic individual-based theory of population dynamics and to explore the relationship between these results and ours, as has previously been considered for directly transmitted diseases \citep{pakkanen_unifying_2023}. These stochastic models may also offer a more natural framework for inferring their parameters (including $R_t$) than our deterministic model.

The key advance of our research is to derive renewal equations from first principles that can be used to model host-vector disease outbreaks. Since renewal equations form the basis of methods for inferring characteristics of pathogen transmission (e.g., $R(t)$) in real-time during outbreaks, we hope that this research is an initial step towards enabling public health policy advisors to track pathogen transmission more effectively during future outbreaks of a wide range of mosquito-borne diseases.

\newpage
\subsection*{Funding}
C.M is supported by a studentship (EP/T517811/1) from the UK’s Engineering and Physical Sciences Research Council. T.A. wishes to acknowledge the financial support of the Kuwait Foundation for the Advancement of Sciences (KFAS). K.V.P. acknowledges funding from the MRC Centre for Global Infectious Disease Analysis (Reference No. MR/X020258/1) funded by the UK Medical Research Council. This UK-funded grant is carried out in the frame of the Global Health EDCTP3 Joint Undertaking. M.U.G.K. acknowledges funding from The Rockefeller Foundation (PC-2022-POP-005), Google.org, the Oxford Martin School Programmes in Pandemic Genomics \& Digital Pandemic Preparedness, European Union's Horizon Europe programme projects MOOD (No. 874850) and E4Warning (No. 101086640), the John Fell Fund, a Branco Weiss Fellowship and Wellcome Trust grants 225288/Z/22/Z, 226052/Z/22/Z \& 228186/Z/23/Z, United Kingdom Research and Innovation (No. APP8583) and the Medical Research Foundation (MRF-RG-ICCH-2022-100069). The contents of this publication are the sole responsibility of the authors and do not necessarily reflect the views of the European Commission or the other funders. C.A.D. is supported by the UK National Institute for Health Research Health Protection Research Unit (NIHR HPRU) in Emerging and Zoonotic Infections in partnership with Public Health England (PHE), Grant Number: HPRU200907.

\subsection*{Data and code availability}
 All code and data used in this article are available at  \url{https://github.com/cathalmills/Rt_dengue}.

\bibliographystyle{abbrvnat}
\bibliography{references}

\begin{thebibliography}{46}
\providecommand{\natexlab}[1]{#1}
\providecommand{\url}[1]{\texttt{#1}}
\expandafter\ifx\csname urlstyle\endcsname\relax
  \providecommand{\doi}[1]{doi: #1}\else
  \providecommand{\doi}{doi: \begingroup \urlstyle{rm}\Url}\fi

\bibitem[Abbott et~al.(2020)Abbott, Hellewell, Thompson, Sherratt, Gibbs, Bosse, Munday, Meakin, Doughty, Chun, Chan, Finger, Campbell, Endo, Pearson, Gimma, Russell, {CMMID COVID modelling group}, Flasche, Kucharski, Eggo, and Funk]{abbott_estimating_2020}
S.~Abbott, J.~Hellewell, R.~N. Thompson, K.~Sherratt, H.~P. Gibbs, N.~I. Bosse, J.~D. Munday, S.~Meakin, E.~L. Doughty, J.~Y. Chun, Y.-W.~D. Chan, F.~Finger, P.~Campbell, A.~Endo, C.~A.~B. Pearson, A.~Gimma, T.~Russell, {CMMID COVID modelling group}, S.~Flasche, A.~J. Kucharski, R.~M. Eggo, and S.~Funk.
\newblock Estimating the time-varying reproduction number of {SARS}-{CoV}-2 using national and subnational case counts.
\newblock \emph{Wellcome Open Research}, 5:\penalty0 112, Dec. 2020.
\newblock ISSN 2398-502X.
\newblock \doi{10.12688/wellcomeopenres.16006.2}.
\newblock URL \url{https://wellcomeopenresearch.org/articles/5-112/v2}.

\bibitem[Brady et~al.(2013)Brady, Johansson, Guerra, Bhatt, Golding, Pigott, Delatte, Grech, Leisnham, Maciel-de Freitas, Styer, Smith, Scott, Gething, and Hay]{brady_modelling_2013}
O.~J. Brady, M.~A. Johansson, C.~A. Guerra, S.~Bhatt, N.~Golding, D.~M. Pigott, H.~Delatte, M.~G. Grech, P.~T. Leisnham, R.~Maciel-de Freitas, L.~M. Styer, D.~L. Smith, T.~W. Scott, P.~W. Gething, and S.~I. Hay.
\newblock Modelling adult \textit{{Aedes} aegypti} and \textit{{Aedes} albopictus} survival at different temperatures in laboratory and field settings.
\newblock \emph{Parasites \& Vectors}, 6\penalty0 (1):\penalty0 351, Dec. 2013.
\newblock ISSN 1756-3305.
\newblock \doi{10.1186/1756-3305-6-351}.
\newblock URL \url{https://parasitesandvectors.biomedcentral.com/articles/10.1186/1756-3305-6-351}.

\bibitem[Breda et~al.(2012)Breda, Diekmann, De~Graaf, Pugliese, and Vermiglio]{breda_formulation_2012}
D.~Breda, O.~Diekmann, W.~F. De~Graaf, A.~Pugliese, and R.~Vermiglio.
\newblock On the formulation of epidemic models (an appraisal of {Kermack} and {McKendrick}).
\newblock \emph{Journal of Biological Dynamics}, 6:\penalty0 103--117, Sept. 2012.
\newblock ISSN 1751-3758, 1751-3766.
\newblock \doi{10.1080/17513758.2012.716454}.
\newblock URL \url{http://www.tandfonline.com/doi/abs/10.1080/17513758.2012.716454}.

\bibitem[Brouwer(2022)]{brouwer_why_2022}
A.~F. Brouwer.
\newblock Why the {Spectral} {Radius}? {An} intuition-building introduction to the basic reproduction number.
\newblock \emph{Bulletin of Mathematical Biology}, 84\penalty0 (9):\penalty0 96, Sept. 2022.
\newblock ISSN 0092-8240, 1522-9602.
\newblock \doi{10.1007/s11538-022-01057-9}.
\newblock URL \url{https://link.springer.com/10.1007/s11538-022-01057-9}.

\bibitem[Champredon et~al.(2018)Champredon, Dushoff, and Earn]{champredon_equivalence_2018}
D.~Champredon, J.~Dushoff, and D.~J.~D. Earn.
\newblock Equivalence of the {Erlang}-{Distributed} {SEIR} {Epidemic} {Model} and the {Renewal} {Equation}.
\newblock \emph{SIAM Journal on Applied Mathematics}, 78\penalty0 (6):\penalty0 3258--3278, Jan. 2018.
\newblock ISSN 0036-1399, 1095-712X.
\newblock \doi{10.1137/18M1186411}.
\newblock URL \url{https://epubs.siam.org/doi/10.1137/18M1186411}.

\bibitem[Chan and Johansson(2012)]{chan_incubation_2012}
M.~Chan and M.~A. Johansson.
\newblock The incubation periods of dengue viruses.
\newblock \emph{PLOS ONE}, 7\penalty0 (11):\penalty0 e50972, Nov. 2012.
\newblock ISSN 1932-6203.
\newblock \doi{10.1371/journal.pone.0050972}.
\newblock URL \url{https://dx.plos.org/10.1371/journal.pone.0050972}.

\bibitem[Chen et~al.(2024)Chen, Xu, Wang, Liang, Li, Lourenço, Yang, Lin, Wang, Zhao, Cazelles, Song, Liu, Wang, Brady, Cauchemez, and Tian]{chen_indian_2024}
Y.~Chen, Y.~Xu, L.~Wang, Y.~Liang, N.~Li, J.~Lourenço, Y.~Yang, Q.~Lin, L.~Wang, H.~Zhao, B.~Cazelles, H.~Song, Z.~Liu, Z.~Wang, O.~J. Brady, S.~Cauchemez, and H.~Tian.
\newblock Indian {Ocean} temperature anomalies predict long-term global dengue trends.
\newblock \emph{Science}, 384\penalty0 (6696):\penalty0 639--646, May 2024.
\newblock ISSN 0036-8075, 1095-9203.
\newblock \doi{10.1126/science.adj4427}.
\newblock URL \url{https://www.science.org/doi/10.1126/science.adj4427}.

\bibitem[Codeço et~al.(2018)Codeço, Villela, and Coelho]{codeco_estimating_2018}
C.~T. Codeço, D.~A. Villela, and F.~C. Coelho.
\newblock Estimating the effective reproduction number of dengue considering temperature-dependent generation intervals.
\newblock \emph{Epidemics}, 25:\penalty0 101--111, Dec. 2018.
\newblock ISSN 17554365.
\newblock \doi{10.1016/j.epidem.2018.05.011}.
\newblock URL \url{https://linkinghub.elsevier.com/retrieve/pii/S1755436517300907}.

\bibitem[Cori et~al.(2013)Cori, Ferguson, Fraser, and Cauchemez]{cori_new_2013}
A.~Cori, N.~M. Ferguson, C.~Fraser, and S.~Cauchemez.
\newblock A new framework and software to estimate time-varying reproduction numbers during epidemics.
\newblock \emph{American Journal of Epidemiology}, 178\penalty0 (9):\penalty0 1505--1512, Nov. 2013.
\newblock ISSN 1476-6256, 0002-9262.
\newblock \doi{10.1093/aje/kwt133}.
\newblock URL \url{https://academic.oup.com/aje/article-lookup/doi/10.1093/aje/kwt133}.

\bibitem[Diekmann and Heesterbeek(2000)]{diekmann_mathematical_2000}
O.~Diekmann and J.~A.~P. Heesterbeek.
\newblock \emph{Mathematical epidemiology of infectious diseases: model building, analysis and interpretation}.
\newblock Wiley series in mathematical and computational biology. Wiley, Chichester Weinheim, 2000.
\newblock ISBN 978-0-471-98682-9 978-0-471-49241-2.

\bibitem[Fraser(2007)]{fraser_estimating_2007}
C.~Fraser.
\newblock Estimating individual and household reproduction numbers in an emerging epidemic.
\newblock \emph{PLOS ONE}, 2\penalty0 (8):\penalty0 e758, Aug. 2007.
\newblock ISSN 1932-6203.
\newblock \doi{10.1371/journal.pone.0000758}.
\newblock URL \url{https://dx.plos.org/10.1371/journal.pone.0000758}.

\bibitem[Funk et~al.(2016)Funk, Kucharski, Camacho, Eggo, Yakob, Murray, and Edmunds]{funk_comparative_2016}
S.~Funk, A.~J. Kucharski, A.~Camacho, R.~M. Eggo, L.~Yakob, L.~M. Murray, and W.~J. Edmunds.
\newblock Comparative analysis of dengue and {Zika} outbreaks reveals differences by setting and virus.
\newblock \emph{PLOS Neglected Tropical Diseases}, 10\penalty0 (12):\penalty0 e0005173, Dec. 2016.
\newblock ISSN 1935-2735.
\newblock \doi{10.1371/journal.pntd.0005173}.
\newblock URL \url{https://dx.plos.org/10.1371/journal.pntd.0005173}.

\bibitem[Gostic et~al.(2020)Gostic, McGough, Baskerville, Abbott, Joshi, Tedijanto, Kahn, Niehus, Hay, De~Salazar, Hellewell, Meakin, Munday, Bosse, Sherrat, Thompson, White, Huisman, Scire, Bonhoeffer, Stadler, Wallinga, Funk, Lipsitch, and Cobey]{gostic_practical_2020}
K.~M. Gostic, L.~McGough, E.~B. Baskerville, S.~Abbott, K.~Joshi, C.~Tedijanto, R.~Kahn, R.~Niehus, J.~A. Hay, P.~M. De~Salazar, J.~Hellewell, S.~Meakin, J.~D. Munday, N.~I. Bosse, K.~Sherrat, R.~N. Thompson, L.~F. White, J.~S. Huisman, J.~Scire, S.~Bonhoeffer, T.~Stadler, J.~Wallinga, S.~Funk, M.~Lipsitch, and S.~Cobey.
\newblock Practical considerations for measuring the effective reproductive number, {R}.
\newblock \emph{PLOS Computational Biology}, 16\penalty0 (12):\penalty0 e1008409, Dec. 2020.
\newblock ISSN 1553-7358.
\newblock \doi{10.1371/journal.pcbi.1008409}.
\newblock URL \url{https://dx.plos.org/10.1371/journal.pcbi.1008409}.

\bibitem[Hersbach et~al.(2020)Hersbach, Bell, Berrisford, Hirahara, Horányi, Muñoz‐Sabater, Nicolas, Peubey, Radu, Schepers, Simmons, Soci, Abdalla, Abellan, Balsamo, Bechtold, Biavati, Bidlot, Bonavita, De~Chiara, Dahlgren, Dee, Diamantakis, Dragani, Flemming, Forbes, Fuentes, Geer, Haimberger, Healy, Hogan, Hólm, Janisková, Keeley, Laloyaux, Lopez, Lupu, Radnoti, De~Rosnay, Rozum, Vamborg, Villaume, and Thépaut]{hersbach_era5_2020}
H.~Hersbach, B.~Bell, P.~Berrisford, S.~Hirahara, A.~Horányi, J.~Muñoz‐Sabater, J.~Nicolas, C.~Peubey, R.~Radu, D.~Schepers, A.~Simmons, C.~Soci, S.~Abdalla, X.~Abellan, G.~Balsamo, P.~Bechtold, G.~Biavati, J.~Bidlot, M.~Bonavita, G.~De~Chiara, P.~Dahlgren, D.~Dee, M.~Diamantakis, R.~Dragani, J.~Flemming, R.~Forbes, M.~Fuentes, A.~Geer, L.~Haimberger, S.~Healy, R.~J. Hogan, E.~Hólm, M.~Janisková, S.~Keeley, P.~Laloyaux, P.~Lopez, C.~Lupu, G.~Radnoti, P.~De~Rosnay, I.~Rozum, F.~Vamborg, S.~Villaume, and J.~Thépaut.
\newblock The {ERA5} global reanalysis.
\newblock \emph{Quarterly Journal of the Royal Meteorological Society}, 146\penalty0 (730):\penalty0 1999--2049, July 2020.
\newblock ISSN 0035-9009, 1477-870X.
\newblock \doi{10.1002/qj.3803}.
\newblock URL \url{https://rmets.onlinelibrary.wiley.com/doi/10.1002/qj.3803}.

\bibitem[Huber et~al.(2018)Huber, Childs, Caldwell, and Mordecai]{huber_seasonal_2018}
J.~H. Huber, M.~L. Childs, J.~M. Caldwell, and E.~A. Mordecai.
\newblock Seasonal temperature variation influences climate suitability for dengue, chikungunya, and {Zika} transmission.
\newblock \emph{PLOS Neglected Tropical Diseases}, 12\penalty0 (5):\penalty0 e0006451, May 2018.
\newblock ISSN 1935-2735.
\newblock \doi{10.1371/journal.pntd.0006451}.
\newblock URL \url{https://dx.plos.org/10.1371/journal.pntd.0006451}.

\bibitem[Inaba and Nishiura(2008)]{inaba_state-reproduction_2008}
H.~Inaba and H.~Nishiura.
\newblock The state-reproduction number for a multistate class age structured epidemic system and its application to the asymptomatic transmission model.
\newblock \emph{Mathematical Biosciences}, 216\penalty0 (1):\penalty0 77--89, Nov. 2008.
\newblock ISSN 00255564.
\newblock \doi{10.1016/j.mbs.2008.08.005}.
\newblock URL \url{https://linkinghub.elsevier.com/retrieve/pii/S0025556408001296}.

\bibitem[Keeling et~al.(2008)Keeling, Keeling, and Rohani]{keeling_modeling_2008}
M.~J. Keeling, M.~J. Keeling, and P.~Rohani.
\newblock \emph{Modeling infectious diseases in humans and animals}.
\newblock Princeton Univ. Press, Princeton, NJ, 2008.
\newblock ISBN 978-0-691-11617-4.

\bibitem[Kermack and McKendrick(1927)]{kermack_contribution_1927}
W.~O. Kermack and A.~G. McKendrick.
\newblock A contribution to the mathematical theory of epidemics.
\newblock \emph{Proceedings of the Royal Society of London. Series A}, 115\penalty0 (772):\penalty0 700--721, 1927.
\newblock Publisher: The Royal Society London.

\bibitem[Keyfitz and Keyfitz(1997)]{keyfitz_mckendrick_1997}
B.~Keyfitz and N.~Keyfitz.
\newblock The {McKendrick} partial differential equation and its uses in epidemiology and population study.
\newblock \emph{Mathematical and Computer Modelling}, 26\penalty0 (6):\penalty0 1--9, Sept. 1997.
\newblock ISSN 08957177.
\newblock \doi{10.1016/S0895-7177(97)00165-9}.
\newblock URL \url{https://linkinghub.elsevier.com/retrieve/pii/S0895717797001659}.

\bibitem[Lambert et~al.(2022)Lambert, North, and Godfray]{lambert_meta-analysis_2022}
B.~Lambert, A.~North, and H.~C.~J. Godfray.
\newblock A meta-analysis of longevity estimates of mosquito {Vectors} of disease, May 2022.
\newblock URL \url{http://biorxiv.org/lookup/doi/10.1101/2022.05.30.494059}.

\bibitem[McKendrick(1925)]{mckendrick_applications_1925}
A.~G. McKendrick.
\newblock Applications of mathematics to {Medical} problems.
\newblock \emph{Proceedings of the Edinburgh Mathematical Society}, 44:\penalty0 98--130, Feb. 1925.
\newblock ISSN 0013-0915, 1464-3839.
\newblock \doi{10.1017/S0013091500034428}.
\newblock URL \url{https://www.cambridge.org/core/product/identifier/S0013091500034428/type/journal_article}.

\bibitem[Metz and Diekmann(1986)]{metz_dynamics_1986}
J.~A.~J. Metz and O.~Diekmann.
\newblock \emph{The dynamics of physiologically structured populations}.
\newblock Number~68 in Lecture notes in biomathematics. Springer-Verl, Berlin, 1986.
\newblock ISBN 978-0-387-16786-2 978-3-540-16786-0.

\bibitem[Monteiro et~al.(2019)Monteiro, De~Souza, Amaral, De~Lima, De~Araújo, Ramalho, Martins, Colares, De~Góes~Cavalcanti, and Lima]{monteiro_dengue_2019}
D.~C.~S. Monteiro, N.~V. De~Souza, J.~C. Amaral, K.~B. De~Lima, F.~M.~C. De~Araújo, I.~L.~C. Ramalho, V.~E.~P. Martins, J.~K.~B. Colares, L.~P. De~Góes~Cavalcanti, and D.~M. Lima.
\newblock Dengue: 30 years of cases in an endemic area.
\newblock \emph{Clinics}, 74:\penalty0 e675, 2019.
\newblock ISSN 18075932.
\newblock \doi{10.6061/clinics/2019/e675}.
\newblock URL \url{https://linkinghub.elsevier.com/retrieve/pii/S1807593222006342}.

\bibitem[Mordecai et~al.(2017)Mordecai, Cohen, Evans, Gudapati, Johnson, Lippi, Miazgowicz, Murdock, Rohr, Ryan, Savage, Shocket, Stewart~Ibarra, Thomas, and Weikel]{mordecai_detecting_2017}
E.~A. Mordecai, J.~M. Cohen, M.~V. Evans, P.~Gudapati, L.~R. Johnson, C.~A. Lippi, K.~Miazgowicz, C.~C. Murdock, J.~R. Rohr, S.~J. Ryan, V.~Savage, M.~S. Shocket, A.~Stewart~Ibarra, M.~B. Thomas, and D.~P. Weikel.
\newblock Detecting the impact of temperature on transmission of {Zika}, dengue, and chikungunya using mechanistic models.
\newblock \emph{PLOS Neglected Tropical Diseases}, 11\penalty0 (4):\penalty0 e0005568, Apr. 2017.
\newblock ISSN 1935-2735.
\newblock \doi{10.1371/journal.pntd.0005568}.
\newblock URL \url{https://dx.plos.org/10.1371/journal.pntd.0005568}.

\bibitem[Mordecai et~al.(2019)Mordecai, Caldwell, Grossman, Lippi, Johnson, Neira, Rohr, Ryan, Savage, Shocket, Sippy, Stewart~Ibarra, Thomas, and Villena]{mordecai_thermal_2019}
E.~A. Mordecai, J.~M. Caldwell, M.~K. Grossman, C.~A. Lippi, L.~R. Johnson, M.~Neira, J.~R. Rohr, S.~J. Ryan, V.~Savage, M.~S. Shocket, R.~Sippy, A.~M. Stewart~Ibarra, M.~B. Thomas, and O.~Villena.
\newblock Thermal biology of mosquito‐borne disease.
\newblock \emph{Ecology Letters}, 22\penalty0 (10):\penalty0 1690--1708, Oct. 2019.
\newblock ISSN 1461-023X, 1461-0248.
\newblock \doi{10.1111/ele.13335}.
\newblock URL \url{https://onlinelibrary.wiley.com/doi/10.1111/ele.13335}.

\bibitem[Murray(2002)]{murray_mathematical_2002}
J.~D. Murray.
\newblock \emph{Mathematical biology}.
\newblock Number 17-18 in Interdisciplinary applied mathematics. Springer, New York [etc.], 3rd ed edition, 2002.
\newblock ISBN 978-0-387-95223-9 978-0-387-95228-4.

\bibitem[Nishiura and Halstead(2007)]{nishiura_natural_2007}
H.~Nishiura and S.~Halstead.
\newblock Natural history of dengue virus ({DENV})–1 and {DENV}‐4 infections: reanalysis of classic studies.
\newblock \emph{The Journal of Infectious Diseases}, 195\penalty0 (7):\penalty0 1007--1013, Apr. 2007.
\newblock ISSN 0022-1899, 1537-6613.
\newblock \doi{10.1086/511825}.
\newblock URL \url{https://academic.oup.com/jid/article-lookup/doi/10.1086/511825}.

\bibitem[Ogi-Gittins et~al.(2023)Ogi-Gittins, Hart, Song, Nash, Polonsky, Cori, Hill, and Thompson]{ogi-gittins_simulation-based_2023}
I.~Ogi-Gittins, W.~Hart, J.~Song, R.~Nash, J.~Polonsky, A.~Cori, E.~Hill, and R.~Thompson.
\newblock A simulation-based approach for estimating the time-dependent reproduction number from temporally aggregated disease incidence time series data, Sept. 2023.
\newblock URL \url{http://medrxiv.org/lookup/doi/10.1101/2023.09.13.23295471}.

\bibitem[Pakkanen et~al.(2023)Pakkanen, Miscouridou, Penn, Whittaker, Berah, Mishra, Mellan, and Bhatt]{pakkanen_unifying_2023}
M.~S. Pakkanen, X.~Miscouridou, M.~J. Penn, C.~Whittaker, T.~Berah, S.~Mishra, T.~A. Mellan, and S.~Bhatt.
\newblock Unifying incidence and prevalence under a time-varying general branching process.
\newblock \emph{Journal of Mathematical Biology}, 87\penalty0 (2):\penalty0 35, Aug. 2023.
\newblock ISSN 0303-6812, 1432-1416.
\newblock \doi{10.1007/s00285-023-01958-w}.
\newblock URL \url{https://link.springer.com/10.1007/s00285-023-01958-w}.

\bibitem[Parag(2021)]{parag_improved_2021}
K.~V. Parag.
\newblock Improved estimation of time-varying reproduction numbers at low case incidence and between epidemic waves.
\newblock \emph{PLOS Computational Biology}, 17\penalty0 (9):\penalty0 e1009347, Sept. 2021.
\newblock ISSN 1553-7358.
\newblock \doi{10.1371/journal.pcbi.1009347}.
\newblock URL \url{https://dx.plos.org/10.1371/journal.pcbi.1009347}.

\bibitem[Parag et~al.(2022)Parag, Thompson, and Donnelly]{parag_are_2022}
K.~V. Parag, R.~N. Thompson, and C.~A. Donnelly.
\newblock Are epidemic growth rates more informative than reproduction numbers?
\newblock \emph{Journal of the Royal Statistical Society Series A: Statistics in Society}, 185:\penalty0 S5--S15, Nov. 2022.
\newblock ISSN 0964-1998, 1467-985X.
\newblock \doi{10.1111/rssa.12867}.
\newblock URL \url{https://academic.oup.com/jrsssa/article/185/Supplement_1/S5/7069470}.

\bibitem[Parag et~al.(2023)Parag, Cowling, and Lambert]{parag_angular_2023}
K.~V. Parag, B.~J. Cowling, and B.~Lambert.
\newblock Angular reproduction numbers improve estimates of transmissibility when disease generation times are misspecified or time-varying.
\newblock \emph{Proceedings of the Royal Society B: Biological Sciences}, 290\penalty0 (2007):\penalty0 20231664, Sept. 2023.
\newblock ISSN 0962-8452, 1471-2954.
\newblock \doi{10.1098/rspb.2023.1664}.
\newblock URL \url{https://royalsocietypublishing.org/doi/10.1098/rspb.2023.1664}.

\bibitem[Park et~al.(2022)Park, Bolker, Funk, Metcalf, Weitz, Grenfell, and Dushoff]{park_importance_2022}
S.~W. Park, B.~M. Bolker, S.~Funk, C.~J.~E. Metcalf, J.~S. Weitz, B.~T. Grenfell, and J.~Dushoff.
\newblock The importance of the generation interval in investigating dynamics and control of new {SARS}-{CoV}-2 variants.
\newblock \emph{Journal of The Royal Society Interface}, 19\penalty0 (191):\penalty0 20220173, June 2022.
\newblock ISSN 1742-5662.
\newblock \doi{10.1098/rsif.2022.0173}.
\newblock URL \url{https://royalsocietypublishing.org/doi/10.1098/rsif.2022.0173}.

\bibitem[Park et~al.(2024)Park, Akhmetzhanov, Charniga, Cori, Davies, Dushoff, Funk, Gostic, Grenfell, Linton, Lipsitch, Lison, Overton, Ward, and Abbott]{park_estimating_2024}
S.~W. Park, A.~R. Akhmetzhanov, K.~Charniga, A.~Cori, N.~G. Davies, J.~Dushoff, S.~Funk, K.~Gostic, B.~Grenfell, N.~M. Linton, M.~Lipsitch, A.~Lison, C.~E. Overton, T.~Ward, and S.~Abbott.
\newblock Estimating epidemiological delay distributions for infectious diseases, Jan. 2024.
\newblock URL \url{http://medrxiv.org/lookup/doi/10.1101/2024.01.12.24301247}.

\bibitem[Romeo-Aznar et~al.(2024)Romeo-Aznar, Telle, Santos-Vega, Paul, and Pascual]{romeo-aznar_crowded_2024}
V.~Romeo-Aznar, O.~Telle, M.~Santos-Vega, R.~Paul, and M.~Pascual.
\newblock Crowded and warmer: {Unequal} dengue risk at high spatial resolution across a megacity of {India}.
\newblock \emph{PLOS Climate}, 3\penalty0 (3):\penalty0 e0000240, Mar. 2024.
\newblock ISSN 2767-3200.
\newblock \doi{10.1371/journal.pclm.0000240}.
\newblock URL \url{https://dx.plos.org/10.1371/journal.pclm.0000240}.

\bibitem[Ross(1910)]{ross_prevention_1910}
R.~Ross.
\newblock \emph{The {Prevention} of malaria}.
\newblock New York, E.P. Dutton \& company, 1910.
\newblock URL \url{https://books.google.com/books?id=0jRaWNX--s0C}.

\bibitem[Shocket et~al.(2020)Shocket, Verwillow, Numazu, Slamani, Cohen, El~Moustaid, Rohr, Johnson, and Mordecai]{shocket_transmission_2020}
M.~S. Shocket, A.~B. Verwillow, M.~G. Numazu, H.~Slamani, J.~M. Cohen, F.~El~Moustaid, J.~Rohr, L.~R. Johnson, and E.~A. Mordecai.
\newblock Transmission of {West} {Nile} and five other temperate mosquito-borne viruses peaks at temperatures between 23°{C} and 26°{C}.
\newblock \emph{eLife}, 9:\penalty0 e58511, Sept. 2020.
\newblock ISSN 2050-084X.
\newblock \doi{10.7554/eLife.58511}.
\newblock URL \url{https://elifesciences.org/articles/58511}.

\bibitem[Singapore(2024)]{meteorological_service_singapore_historical_2024}
M.~S. Singapore.
\newblock Historical {Daily} {Records}, Aug. 2024.
\newblock URL \url{https://www.weather.gov.sg/climate-historical-daily/}.

\bibitem[Siraj et~al.(2017)Siraj, Oidtman, Huber, Kraemer, Brady, Johansson, and Perkins]{siraj_temperature_2017}
A.~S. Siraj, R.~J. Oidtman, J.~H. Huber, M.~U.~G. Kraemer, O.~J. Brady, M.~A. Johansson, and T.~A. Perkins.
\newblock Temperature modulates dengue virus epidemic growth rates through its effects on reproduction numbers and generation intervals.
\newblock \emph{PLOS Neglected Tropical Diseases}, 11\penalty0 (7):\penalty0 e0005797, July 2017.
\newblock ISSN 1935-2735.
\newblock \doi{10.1371/journal.pntd.0005797}.
\newblock URL \url{https://dx.plos.org/10.1371/journal.pntd.0005797}.

\bibitem[Smith et~al.(2012)Smith, Battle, Hay, Barker, Scott, and McKenzie]{smith_ross_2012}
D.~L. Smith, K.~E. Battle, S.~I. Hay, C.~M. Barker, T.~W. Scott, and F.~E. McKenzie.
\newblock Ross, {Macdonald}, and a theory for the dynamics and control of mosquito-transmitted pathogens.
\newblock \emph{PLOS Pathogens}, 8\penalty0 (4):\penalty0 e1002588, Apr. 2012.
\newblock ISSN 1553-7374.
\newblock \doi{10.1371/journal.ppat.1002588}.
\newblock URL \url{https://dx.plos.org/10.1371/journal.ppat.1002588}.

\bibitem[Svensson(2007)]{svensson_note_2007}
A.~Svensson.
\newblock A note on generation times in epidemic models.
\newblock \emph{Mathematical Biosciences}, 208\penalty0 (1):\penalty0 300--311, July 2007.
\newblock ISSN 00255564.
\newblock \doi{10.1016/j.mbs.2006.10.010}.
\newblock URL \url{https://linkinghub.elsevier.com/retrieve/pii/S0025556406002094}.

\bibitem[Tesla et~al.(2018)Tesla, Demakovsky, Mordecai, Ryan, Bonds, Ngonghala, Brindley, and Murdock]{tesla_temperature_2018}
B.~Tesla, L.~R. Demakovsky, E.~A. Mordecai, S.~J. Ryan, M.~H. Bonds, C.~N. Ngonghala, M.~A. Brindley, and C.~C. Murdock.
\newblock Temperature drives {Zika} virus transmission: evidence from empirical and mathematical models.
\newblock \emph{Proceedings of the Royal Society B: Biological Sciences}, 285\penalty0 (1884):\penalty0 20180795, Aug. 2018.
\newblock ISSN 0962-8452, 1471-2954.
\newblock \doi{10.1098/rspb.2018.0795}.
\newblock URL \url{https://royalsocietypublishing.org/doi/10.1098/rspb.2018.0795}.

\bibitem[Thompson et~al.(2019)Thompson, Stockwin, Van~Gaalen, Polonsky, Kamvar, Demarsh, Dahlqwist, Li, Miguel, Jombart, Lessler, Cauchemez, and Cori]{thompson_improved_2019}
R.~Thompson, J.~Stockwin, R.~Van~Gaalen, J.~Polonsky, Z.~Kamvar, P.~Demarsh, E.~Dahlqwist, S.~Li, E.~Miguel, T.~Jombart, J.~Lessler, S.~Cauchemez, and A.~Cori.
\newblock Improved inference of time-varying reproduction numbers during infectious disease outbreaks.
\newblock \emph{Epidemics}, 29:\penalty0 100356, Dec. 2019.
\newblock ISSN 17554365.
\newblock \doi{10.1016/j.epidem.2019.100356}.
\newblock URL \url{https://linkinghub.elsevier.com/retrieve/pii/S1755436519300350}.

\bibitem[Vegvari et~al.(2022)Vegvari, Abbott, Ball, Brooks-Pollock, Challen, Collyer, Dangerfield, Gog, Gostic, Heffernan, Hollingsworth, Isham, Kenah, Mollison, Panovska-Griffiths, Pellis, Roberts, Scalia~Tomba, Thompson, and Trapman]{vegvari_commentary_2022}
C.~Vegvari, S.~Abbott, F.~Ball, E.~Brooks-Pollock, R.~Challen, B.~S. Collyer, C.~Dangerfield, J.~R. Gog, K.~M. Gostic, J.~M. Heffernan, T.~D. Hollingsworth, V.~Isham, E.~Kenah, D.~Mollison, J.~Panovska-Griffiths, L.~Pellis, M.~G. Roberts, G.~Scalia~Tomba, R.~N. Thompson, and P.~Trapman.
\newblock Commentary on the use of the reproduction number \textit{{R}} during the {COVID}-19 pandemic.
\newblock \emph{Statistical Methods in Medical Research}, 31\penalty0 (9):\penalty0 1675--1685, Sept. 2022.
\newblock ISSN 0962-2802, 1477-0334.
\newblock \doi{10.1177/09622802211037079}.
\newblock URL \url{http://journals.sagepub.com/doi/10.1177/09622802211037079}.

\bibitem[Wallinga and Lipsitch(2007)]{wallinga_how_2007}
J.~Wallinga and M.~Lipsitch.
\newblock How generation intervals shape the relationship between growth rates and reproductive numbers.
\newblock \emph{Proceedings of the Royal Society B: Biological Sciences}, 274\penalty0 (1609):\penalty0 599--604, Feb. 2007.
\newblock ISSN 0962-8452, 1471-2954.
\newblock \doi{10.1098/rspb.2006.3754}.
\newblock URL \url{https://royalsocietypublishing.org/doi/10.1098/rspb.2006.3754}.

\bibitem[Yang et~al.(2009)Yang, Macoris, Galvani, Andrighetti, and Wanderley]{yang_assessing_2009}
H.~M. Yang, M.~L.~G. Macoris, K.~C. Galvani, M.~T.~M. Andrighetti, and D.~M.~V. Wanderley.
\newblock Assessing the effects of temperature on the population of \textit{{Aedes} aegypti} , the vector of dengue.
\newblock \emph{Epidemiology and Infection}, 137\penalty0 (8):\penalty0 1188--1202, Aug. 2009.
\newblock ISSN 0950-2688, 1469-4409.
\newblock \doi{10.1017/S0950268809002040}.
\newblock URL \url{https://www.cambridge.org/core/product/identifier/S0950268809002040/type/journal_article}.

\end{thebibliography}
\newpage
\begin{appendix}
\renewcommand{\thefigure}{SI~\arabic{figure}}
\setcounter{figure}{0}
\section{Renewal equations, generation times, and McKendrick–von Foerster equations}
The epidemic renewal equation framework forms the backbone of many popular methods for inferring transmissibility throughout the course of an infectious disease outbreak \citep{fraser_estimating_2007, cori_new_2013,thompson_improved_2019,abbott_estimating_2020, parag_improved_2021}. Mathematically, the renewal equation framework (in continuous time) for directly transmitted human-to-human infectious diseases can be written as:
\begin{align}\label{eq:appendix_ubiquituous_renewal}
i(t) &= R(t) \int_{0}^{\infty} i(t-\tau) w(\tau)d\tau
\end{align}
where the instantaneous reproduction number $R(t)$ is the expected number of secondary human infections generated per primary human infection if conditions were to remain the same as at time $t$, and $i(t-\tau)$ is the number of human infections that occurred at time $t-\tau$ which is weighted by the probability density of the generation time $w(\tau)$ (by convention, $w(\tau) = 0$ for $\tau < 0$ and $\int_0^\infty w(\tau) d\tau$ = 1). The framework is most often implemented in discrete time to align with the temporal resolution of surveillance data.

Whilst the generation time distribution can be viewed as the distribution of times between primary human infection and secondary transmission, it is also useful to view the generation time distribution as the expected relative contribution to the force of infection by a human who was infected a specific number of time units ago (here, $\tau$ units ago) \citep{breda_formulation_2012, park_importance_2022}. The latter interpretation is useful for our proposed frameworks. It is important to distinguish that the renewal equation, eq. \eqref{eq:appendix_ubiquituous_renewal}, requires the intrinsic generation time distribution (i.e. the time between primary human infection and secondary transmission assuming constant transmission conditions), as opposed to other epidemiological intervals (such as transmission intervals, serial intervals, observed generation times etc).

The generation time is a key (and often overlooked \citep{gostic_practical_2020, park_estimating_2024}) modelling ingredient for any renewal equation framework -- its distribution controls times between a primary human infection and secondary human infection. However, when using renewal equation frameworks, the generation time distribution is commonly assumed to be known/fixed (and not estimated), and is often approximated using the serial interval (the time from symptom onset of the primary case to symptom onset in the secondary case). The use of a time-homogeneous generation time distribution (fixed over calendar time), as in eq. \eqref{eq:appendix_ubiquituous_renewal}, can often be justified (partially) using both statistical (changes to $R(t)$ and $w$ are often not simultaneously identifiable from available surveillance data) and epidemiological (as an intrinsic generation time distribution is time-invariant in the setting of the constant-strength changes which only impact the epidemic strength -- $R(t)$) arguments. Nevertheless, the assumption of a fixed, known generation time distribution means that misspecification of the generation time distribution can be a large source of bias when estimating $R(t)$ and of the true branching process dynamics of an epidemic \citep{parag_are_2022, park_importance_2022, parag_angular_2023, park_estimating_2024}. Given the popularity of $R(t)$ for informing public health policy and decision-making \citep{vegvari_commentary_2022}, the violation of the assumption of time-invariant generation times clearly has practical, real-world consequences. 

Time-varying (also known as instantaneous or time-inhomogeneous) generation time distributions are less commonly used. These allow for the relative contribution of individuals with specific ages (i.e. times since infection) to vary over calendar time, and can be derived as the normalised version of the instantaneous kernel \citep{park_importance_2022}: 
\begin{align*}
    g(t, \tau) := \frac{K(t, \tau)}{R(t)}
\end{align*}
where $K(t, \tau)$ is the instantaneous kernel which describes the rate at which secondary human infections are generated at calendar time $t$ by individuals who were infected $\tau$ time units ago, and $R(t) = \int_{0}^\infty K(t, \tau) d\tau$ is the instantaneous reproduction number which has the same interpretation as in eq. \eqref{eq:ubiquituous_renewal}. 


For vector-borne pathogens such as dengue, the assumption of a fixed generation time distribution over calendar time is highly unrealistic due to mosquito-viral traits (such as the extrinsic incubation period, the rate of mosquito blood feeding, and mosquito density) which have estimated potentially strong temperature dependencies \citep{chan_incubation_2012, mordecai_detecting_2017, mordecai_thermal_2019}. Similarly, the length of time between primary and secondary human infections is likely influenced by the totals of susceptible and infectious vectors and host. So, in addition to the multi-stage nature of the transmission cycle, the ubiquitous renewal equation framework, eq. \eqref{eq:appendix_ubiquituous_renewal}, is not directly applicable for vector-borne diseases. 

Focusing on dengue, recent works (outside renewal equation frameworks) have leveraged empirical data to derive time-varying, temperature-dependent generation time distributions. For instance, the generation time distribution has been modelled as a convolution of temperature-dependent parametric distributions for each of the four stages of the dengue transmission cycle (i.e. from the primary human infection to subsequent human transmission). Standard Ross-Macdonald, temperature-dependent assumptions for mosquito-borne pathogens \citep{ross_prevention_1910, smith_ross_2012} were used to calculate a temperature-driven basic reproduction-number $R_0$ (and exponential growth rate $r_t$) from elements derived from the generation time (such as driven death rates, mosquito-to-human ratios, and the extrinsic incubation period, all of which were determined by temperature) \citep{siraj_temperature_2017}. Similarly, another framework uses time-varying, temperature-dependent generation time distributions and the same empirical data and constituent phases of the transmission cycle, albeit with alternative distributional forms (gamma distributions) for mathematical convenience \citep{codeco_estimating_2018}. The generation time distribution was determined in a piecewise constant, recursive form by assuming temperature varies in discrete steps (weekly here). Then, an instantaneous reproduction number $R(t)$ at time $t$ was derived using the assumptions of \citep{wallinga_how_2007}, the case count data up to time $t$, and the derived time-varying, temperature-driven generation time distribution. Other mechanistic approaches have also regularly deployed extensions of the Ross-Macdonald framework, yielding basic reproduction numbers determined by temperature, as above and in \citep{mordecai_thermal_2019}, and basic reproduction numbers determined by temperature and vector carrying capacity per human \citep{romeo-aznar_crowded_2024}. Nevertheless, across each of the existing methods, there has been a strong dependence on temperature forcing. Furthermore, there has been a lack of focus on a clear mechanistic, mathematical account of how the derived instantaneuous reproduction number $R$ is determined by the calendar-time-varying properties of the human and vector sub-populations that form the stages of the transmission cycle of the vector-borne pathogen. Similarly, there is no clear understanding of what it means to assume a temporally evolving generation time distribution for each constituent phase and sub-population involved in the human-to-human transmission cycle.

\newpage
\section{Exposure-and-infection-age-structured human and infection-age-structured mosquito population}\label{sec:exposed_and_infection_structured_model_2_appendix}
The following section presents additional details of derivations for the framework initially discussed in Section \ref{sec:infection_structured_model_2}. 
\subsection{Modelling framework}
Recall that we had previously the following system of age-structured populations at calendar times $t$, with $E$ and $M$ denoting the density of the infected (and here, also infectious) human and mosquito populations respectively (with corresponding ages $a_E$ and $a_M$):
\begin{align}
    \frac{\partial E}{\partial t}+\frac{\partial E}{\partial a_E} &=0,  \label{appendix_eq:exposed_model_2}\\
  \frac{\partial M}{\partial t}+\frac{\partial M}{\partial a_M} &=-\mu M(t,a_M),  \label{appendix_eq:mosquito_model_2}  
\end{align}
where $\mu>0$ is the (constant) mortality rate in the (infected) mosquito population and we close the system with the following initial and boundary conditions:
\begin{align}
    E(t,0) &= \int_{0}^{\infty} b_{ME}(t,a_M) M(t,a_M) da_M,\label{appendix_eq:conservation_model_2}\\
    E(0,a_E) &= f_E(a_E),\\
    M(t,0) &= \int_{0}^{\infty} b_{EM}(t,a_E) E(t,a_E) da_E,\label{appendix_eq:conservation_model_2_m}\\
    M(0,a_M) &= f_M(a_M) \label{appendix_eq:initial_mosquito_model_2}.
\end{align}
where $f_E(a_E)$ and $f_M(a_M)$ denote the initial densities of infectious human and mosquito populations, and each $b_{AB}(t,a_A) \geq 0$ is a time- and age-dependent birth function that represents the rate at which new individuals in sub-population $B$ are generated at calendar time $t$ by an individual of age $a_A$ in sub-population $A$. As a linear PDE with suitable initial and boundary conditions creates a Cauchy problem, we can now derive formulae for renewal equations, instantaneous reproduction numbers, and time-varying generation time distributions using the method of characteristics (which tries to recreate the solution surface of our systems).

\subsection{Deriving a renewal equation and instantaneous reproduction number} 
Now, we provide additional details of the derivation for our characteristic equation, eq. \eqref{eq:phi_gamma_model_2}. First, considering the characteristics for the mosquito population, we have
\begin{equation}\label{appendix_eq:characteristics}
  a_M=\begin{cases}
    t-t_0, & \text{if $a_M\leq t$},\\
    t+a_0, & \text{if $a_M> t$}.
  \end{cases}
\end{equation}
where $t_0$ is the time at which an individual mosquito of age $0 < a_M < t$ is born, and $a_0$ is the initial age of an individual mosquito who has age $a_M > t$ at time $t$. For each of the two cases given by eq. \eqref{appendix_eq:characteristics}, we can determine the solution:
\begin{equation}
    M(t,a_M) =
    \begin{cases}
    M(t-a_M,0) e^{-\mu a_M}, & \text{if $a_M\leq t$},\\
    f_M(a_M-t) e^{-\mu t}, & \text{if $a_M> t$}.
  \end{cases}
\end{equation}
This means that we can rewrite eq. \eqref{appendix_eq:conservation_model_2} as:
\begin{equation}\label{appendix_eq:renewal_model_2_complex}
    E(t,0) = \int_{0}^{t} b_{ME}(t,a_M) M(t-a_M,0) e^{-\mu a_M} da_M + \int_{t}^{\infty} f_M(a_M-t) e^{-\mu t} da_M,
\end{equation}
where we are mostly concerned with the long-time behaviour of the system, i.e. when $t\gg a_M$ and $f_M(a_M-t) e^{-\mu t}\approx 0$. Then we can neglect the second part of eq. \eqref{appendix_eq:renewal_model_2_complex} and rewrite it as:
\begin{equation}\label{appendix_eq:renewal_model_2_simple}
    E(t,0) = \int_{0}^{t} b_{ME}(t,a_M) M(t-a_M,0) e^{-\mu a_M} da_M.
\end{equation}
We finally use substitute eq. \eqref{appendix_eq:conservation_model_2_m} into eq. \eqref{appendix_eq:renewal_model_2_simple} to yield the renewal equation:
\begin{equation}\label{appendix_eq:renewal_model_2}
     E(t,0) = \int_{0}^{t} \int_{0}^{\infty} b_{ME}(t,a_M)  e^{-\mu a_M}  b_{EM}(t-a_M,a_E) E(t-a_M,a_E) da_E da_M.
\end{equation}
Eq. \eqref{appendix_eq:renewal_model_2} makes explicit the fact that mosquito mortality needs to be taken account of when using renewal equations to model vector-borne diseases.

In the same way as for the mosquito population, we can write the solution for the human population as follows:
\begin{equation}
    E(t,a_E) =
    \begin{cases}
    E(t-a_E,0), & \text{if $a_E\leq t$},\\
    f_E(a_E-t), & \text{if $a_E> t$}.
  \end{cases}
\end{equation}
This means that, focusing on long-term behaviour of the system ((i.e. if $t$ so large ($t \rightarrow \infty$) such that $t>a_E$, then $f_E(a_E - t) = 0$.), we can rewrite eq. \eqref{appendix_eq:renewal_model_2} as
\begin{equation}\label{appendix_eq:renewal_2d_human_expanded_model_2}
E(t,0) = \int_{0}^{t} \int_{0}^{t-a_M} b_{ME}(t,a_M)  e^{-\mu a_M}  b_{EM}(t-a_M,a_E) E(t-a_M-a_E, 0) da_E da_M.
\end{equation}
We now seek a long-term solution of the form: $E(t,a_E) = p(a_E) e^{\gamma t}$. This solution states that the age distribution of exposed individuals is altered by a factor which grows or decays with time depending on whether $\gamma > 0$ or $\gamma < 0$ respectively. We first substitute this into eq. \eqref{appendix_eq:exposed_model_2} to yield:
\begin{equation}
    \gamma p(a_E) e^{\gamma t} + \frac{d p(a_E)}{da_E} e^{\gamma t} = 0,
\end{equation}
which implies
\begin{equation}\label{appendix_eq:model_2_age}
    p(a_E) = p(0) e^{-\gamma a_E}.
\end{equation}
Now substituting the long-term solution and eq. \eqref{appendix_eq:model_2_age} into eq. \eqref{appendix_eq:renewal_model_2} yields:
\begin{equation}
    p(0) e^{\gamma t} = \int_{0}^{t} \int_{0}^{t-a_M} b_{ME}(t,a_M)  e^{-\mu a_M}  b_{EM}(t-a_M,a_E) p(0) e^{\gamma(t-a_M- a_E)}  da_E da_M,
\end{equation}
which upon cancellation of $p(0) e^{\gamma t}$ from both sides gives:
\begin{equation}\label{eq:appendix_implicit_eq}
1 = \int_{0}^{t} \int_{0}^{t - a_M} b_{ME}(t,a_M)  e^{-\mu a_M}  b_{EM}(t-a_M,a_E) e^{-\gamma (a_E+a_M)}  da_E da_M := \phi(\gamma).
\end{equation}
\newpage
\subsection{Relation to existing renewal equation frameworks}\label{appendix:model_2_relation}
We provide here the additional details for how our renewal equation, eq. \eqref{eq:renewal_2d_model_2_overall_gi}, can be related to existing renewal equation frameworks for directly transmitted infectious diseases (where a homogeneous delay distribution over calendar time is typically assumed). 

Denoting $\tau = a_E + a_M$ as the overall delay between primary and secondary human infections, the relation (and compression) of the density for our generation time $g(t, a_M, a_E)$ to a 1-D marginal density $w(t, \tau)$ at time $t$), can be established by i) a Jacobian-based change of variable (to parameterise our integral equation at time $t$ in terms of $\tau$) and ii) $C-1=1$ integration steps (where $C=2$ is the number of model compartments) to marginalise out the first times-between-state variable ($a_E$). 

Focusing on the integral
\begin{equation*}
    \int_{0}^{t} \int_{0}^{t-a_M} g(t,a_M,a_E) E(t-a_M-a_E,0)da_E da_M
\end{equation*} 
from eq. \eqref{eq:renewal_2d_model_2_overall_gi}, we will now use a change of variables to convert $(a_E, a_M) \rightarrow (a_E, \tau)$. Using bijective functions $l(a_E, \tau) = a_M = \tau - a_E$, and $m(a_E, \tau) = a_E$, we have
\begin{equation}\label{appendix_eq:deriving_renewal_2d_model_2_additive_delay_inhomogeneous_gi}
\begin{aligned}
    \int_{0}^{t} \int_{0}^{t-a_M} g(t,a_M,a_E) E(t-a_M-a_E,0)da_E da_M \\
    = \int_{0}^{t} \int_{0}^{\tau} g(t,l(a_E, \tau), m(a_E, \tau)) E(t-l(a_E, \tau) - m(a_E, \tau),0) \left|\frac{\partial(a_E, a_M)}{\partial(a_E, \tau)}\right|da_E d\tau \\
    = \int_{0}^{t} \int_{0}^{\tau} g(t, \tau - a_E, a_E) E(t-(\tau - a_E)-a_E,0) da_E d\tau \\
      = \int_{0}^{t} E(t-\tau,0) \int_{0}^{\tau} g(t, \tau - a_E, a_E) da_E d\tau \\
            = \int_{0}^{t} E(t-\tau,0) \left[\tilde{g}(t, \tau - a_E, a_E)\right]_0^\tau \ d\tau \\
                        = \int_{0}^{t} E(t-\tau,0) w(t,\tau) d\tau
\end{aligned}
\end{equation}
where $ \left|\frac{\partial(a_E, a_M)}{\partial(a_E, \tau)}\right| = 1$ denotes the Jacobian (i.e. the determinant of the Jacobian matrix), $\tilde{g}(t, \tau - a_E, a_E)$ is the antiderivative of $g(t, \tau - a_E, a_E)$ with respect to $a_E$, and $w(t, \tau)$ is the corresponding definite integral defined as  $w(t, \tau):= \left[\tilde{g}(t, \tau - a_E, a_E)\right]_0^\tau$ (which is again evaluated with respect to $a_E$). Note that $g(t,l(a_E, \tau), m(a_E, \tau))$ was rewritten as $g(t, \tau - a_E, a_E)$.

Then, our renewal equation becomes
\begin{equation}\label{appendix_eq:renewal_2d_model_2_additive_delay_inhomogeneous_gi}
     E(t,0) = R(t)\int_{0}^{t}  w(t,\tau) E(t-\tau,0)d\tau.
\end{equation}
where $w(t, \tau):= \left[\tilde{g}(t, \tau - a_E, a_E)\right]_0^\tau = \int_{0}^{\tau} g(t, \tau - a_E, a_E) da_E$.

\newpage
If we additionally assume a homogeneous generation time distribution over calendar time, (i.e. $g(t,a_M,a_E) = g(a_M,a_E) \forall t$, we would have the familiar renewal equation framework: 
\begin{align}\label{appendix_eq:renewal_2d_model_2_additive_delay_homogeneous_gi}
     E(t,0) = R(t)\int_{0}^{t}  w(\tau) E(t-\tau,0)d\tau.
\end{align}
Assuming a homogeneous generation time distribution over calendar time would mean that the ratio defined by
\begin{align}
\label{appendix_eq:model_2_time_varying_gi_meaning}
    g(t,a_M,a_E) &= \frac{K(t,a_M,a_E)}{R(t)}   &= \frac{b_{ME}(t,a_M)  e^{-\mu a_M} b_{EM}(t-a_M,a_E)}{\int_{0}^{t} \int_{0}^{t-a_M} b_{ME}(t,a_M)  e^{-\mu a_M}  b_{EM}(t-a_M,a_E) da_E da_M}, 
\end{align}
is fixed over calendar time ($t$). So, the typically implicit modelling assumption means that the relative contribution (to secondary human infections) of each times-since-compartment-entry pair ($a_M, a_E$) to new current human infections is constant throughout the course of an outbreak. In the current framework, such an assumption means that the expected contribution (to new human infections) of each \textit{individual} pair of times-in-states ($K(t, a_M,a_E)$ scales linearly over calendar time $t$ with the expected \textit{overall} number of new infections per initially infected human ($R(t)$).

The assumption may be unrealistic for a vector-borne disease with a multi-stage transmission cycle such as dengue (e.g. higher temperatures may reduce the extrinsic incubation period \citep{chan_incubation_2012} and/or changes in the host and vector population sizes may impact the distribution of times between generation events). 

Nevertheless, eq. \eqref{appendix_eq:renewal_2d_model_2_additive_delay_homogeneous_gi} formalises the relationship between our framework and the conventional familiar renewal equations (used in e.g. \citep{fraser_estimating_2007, cori_new_2013}). These a special case of our general framework (with $C=2$ compartments). The conventional renewal equations additionally assume a constant generation time distribution over calendar time. Until now, the 1-D generation time distribution described by $w(\tau)$ has been included in frameworks directly as a function of the scalar delay $\tau$ (time from primary human to secondary human infection) where $\tau$ has not been decomposed into the times-since-state-entry variables and their corresponding joint density (as specified by $g$).

\subsection*{Monte Carlo simulation}
Now, we demonstrate how to evaluate the 2-D integral in eq. \eqref{eq:renewal_2d_model_2_overall_gi} via Monte Carlo simulation. The simulation procedure simply requires samples ($a_M, a_E$) from $g(t,a_M,a_E)$, then $\tau^i = a_M^i + a_E^i $. This is equivalent to obtaining samples $\tau$ from $w(t,\tau)$, as defined in eq. \eqref{appendix_eq:deriving_renewal_2d_model_2_additive_delay_inhomogeneous_gi}. This procedure yields
\begin{equation*}
\begin{aligned}
\int_{0}^{t} \int_{0}^{t-a_M} g(t,a_M,a_E) E(t-a_M-a_E,0)da_E da_M &= \mathbb{E}_g[E(t-a_M-a_E,0)] \\ 
&= \mathbb{E}_g[E(t-\tau,0)] \\ 
&\approx \frac{1}{N} \sum_{i=1}^{N} E(t-\tau^i, 0)
\end{aligned}
\end{equation*}where $\mathbb{E}_g$ denotes expectation with respect to the probability density for the (time-varying) generation time $g(t,a_M,a_E)$. We use the same Monte Carlo sampling procedure for our full four-compartment model (see Section \ref{sec:exposed_and_infection_structured_model_4}) when deriving the density for our generation time over calendar time.
\newpage

\subsection{Analytical solutions for $R_0$ and $r$ \label{sec:analytical_appendix}}
To derive analytical solutions for $R(t)$ and $r_t$, we make the simplifying assumption that $b_{EM}(t, a_E) = e^{-\lambda_{E}a_E}$ and $b_{ME}(t, a_M) = e^{-\lambda_{M}a_M}$ which assumes that (for positive rate parameters) the birth processes decline with time since infection. As we have time-independent birth processes, we are actually deriving a basic reproduction number $R_0$ and intrinsic growth rate $r$ where we take 
$t \rightarrow \infty$ in eq. \eqref{eq:phi_gamma_model_2} for the upper integral limits. So, eq. \eqref{eq:phi_gamma_model_2} becomes
\begin{equation}\label{appendix_eq:r_R_model_2_eq_1}
\begin{aligned}
        1 &= \phi(\gamma) \\
        &= \int_{0}^{\infty}  e^{-(\mu + \lambda_M  + \gamma) a_M} \int_{0}^{\infty} e^{-(\lambda_{E} + \gamma) a_E}  da_E da_M  \\
        &=  \frac{1}{\lambda_E + \gamma}  \int_{0}^{\infty}  e^{-(\mu + \lambda_M  + \gamma) a_M}da_M \\
        &=  \frac{1}{\lambda_E + \gamma} \cdot \frac{1}{\lambda_M + \gamma + \mu}  \\
        \end{aligned}
\end{equation}

So, $r$ is the unique real root of eq. \eqref{appendix_eq:r_R_model_2_eq_1}, and our basic reproduction number $R_0$ is the critical threshold for epidemic growth defined as
\begin{equation}
                        R_0:= \phi(0) = \frac{1}{\lambda_E}\cdot\frac{1}{\lambda_M + \mu}
\end{equation}

For understanding the relationship between $R_0$ and $r$, we can use eq. \eqref{eq:R_r_relationship} together with our assumptions above. Then,
\begin{equation}
\begin{aligned}
\frac{1}{R_0} &= \frac{\int_{0}^{\infty}  e^{-(\mu + \lambda_M  + r) a_M} \int_{0}^{\infty} e^{-(\lambda_{E} + r) a_E}  da_E da_M}{\int_{0}^{\infty}  e^{-(\mu + \lambda_M) a_M} \int_{0}^{\infty} e^{-\lambda_{E} a_E}  da_E da_M} \\ 
&= \frac{\lambda_E(\lambda_M + \mu)}{(\lambda_E + r)(\lambda_M + \mu + r)}
                        \end{aligned}
\end{equation}
So, $R_0$ can also be written as a function of $r$ (together with $\lambda_E, \lambda_M$, and $\mu$) as follows:
\begin{equation}\label{eq:R_as_function_of_r}
    R_0 = \frac{(\lambda_E + r)(\lambda_M + \mu + r)}{\lambda_E(\lambda_M + \mu)}
\end{equation}
\section{Exposure-and-infection-age-structured human and infection-age-structured mosquito population}\label{sec:exposed_and_infection_structured_model_3}
\subsection{Modelling framework}
We now describe an intermediate framework (with three compartments) between the frameworks described in Section \ref{sec:infection_structured_model_2} and Section \ref{sec:exposed_and_infection_structured_model_4}. Building on the framework in Section \ref{sec:infection_structured_model_2}, we incorporate separate exposed and infectious human compartments (E and I respectively), which are again structured by times since exposure and infection respectively. The infectious mosquito population is also structured by time since infection. The framework is described by the following conservation equations:
\begin{align}
\frac{\partial E}{\partial t}+\frac{\partial E}{\partial a_E} &=-b_{E I}\left(t, a_{E}\right) E\left(t, a_{E}\right) \label{appendix_eq:exposed_model_three}
\\
\frac{\partial I}{\partial t}+\frac{\partial I}{\partial a_I}&=0 
\label{appendix_eq:infectious_model_three}
\\
\frac{\partial M}{\partial t}+\frac{\partial M}{\partial a_M}&=-\mu M\left(t, a_M\right) \label{appendix_eq:mosquito_model_three}
\end{align}
We close the system with the following initial and boundary conditions:
\begin{align}
E(t, 0) &=\int_0^\infty b_{ME}\left(t, a_M\right) M\left(t, a_M\right) d a_M  \label{appendix_eq:boundary_on_exposed_model_three}\\
E\left(0, a_E\right) &=f_E\left(a_E\right) \\
I(t, 0) &=\int_0^{\infty} b_{E I}\left(t, a_E\right) E\left(t, a_E\right) d a_E \label{appendix_eq:boundary_on_infectious_model_three} \\
I\left(0, a_I\right) &=f_I\left(a_I\right) \\
M(t, 0) &=\int_0^{\infty} b_{IM}\left(t, a_I\right) I\left(t, a_I\right) d a_I \label{appendix_eq:boundary_on_mosquitoes_model_three} \\
M\left(0, a_M\right) &=f_M\left(a_M\right)  \\
\end{align}
where $\mu \geq 0$ is the infectious mosquito death rate (constant across time and ages), $f_E$, $f_I$ and $f_M$ denote the initial densities for the exposed human, infectious human, and infectious mosquito populations respectively, and each $b_{AB}(t,a_A) \geq 0$ is a time- and age-dependent birth function that represents the rate at which new individuals in sub-population $B$ are generated at calendar time $t$ by an individual of age $a_A$ in sub-population $A$.

\subsection{Deriving a renewal equation and instantaneous reproduction number} 

We first consider the characteristics for the mosquito population (for which there is no distinction between exposed or infectious states):
\begin{equation}
    M(t,a_M) =
    \begin{cases}
    M(t-a_M,0) e^{-\mu a_M}, & \text{if $a_M\leq t$},\\
    f_M(a_M-t) e^{-\mu t}, & \text{if $a_M> t$}.
  \end{cases}
\end{equation}
Then, we can rewrite \eqref{appendix_eq:boundary_on_exposed_model_three} as:
\begin{equation}    
\label{appendix_eq:renewal_model_3_complex}
\begin{aligned} 
E(t, 0) = \int_0^t b_{ME}\left(t, a_M\right) M(t-a_M,0) e^{-\mu a_M} d a_M + \\ \int_t^\infty b_{ME}\left(t, a_M\right) f_M(a_M-t) e^{-\mu t} d a_M 
\end{aligned}
\end{equation}
Again, as we are mostly concerned with the long-time behaviour of the system, i.e. when $t\gg a_M$ and $f_E(a_E - t) e^{-\mu t}\approx 0$. Then we can neglect the second part of eq. \eqref{appendix_eq:renewal_model_3_complex} and rewrite it as:
\begin{align} \label{appendix_eq:renewal_model_3_complex_2}
E(t, 0) = \int_0^t b_{ME}\left(t, a_M\right) M(t-a_M,0) e^{-\mu a_M} d a_M 
\end{align}
Now using eq. \eqref{appendix_eq:boundary_on_mosquitoes_model_three}, we have:
\begin{align} \label{appendix_eq:renewal_model_3_complex_3}
E(t, 0) = \int_0^t b_{ME}\left(t, a_M\right) \int_0^{\infty} b_{IM}\left(t-a_M, a_I\right) I\left(t-a_M, a_I\right) d a_I e^{-\mu a_M} d a_M 
\end{align}
Upon writing the characteristics of the infectious sub-population, for each of the two cases, we have the solution:
\begin{equation} \label{appendix_eq:characteristics_cases_infectious_model_three}
    I(t,a_I) =
    \begin{cases}
    I(t - a_I,0), & \text{if $a_I\leq t$},\\
        f_I(a_I - t), & \text{if $a_I > t$},
  \end{cases}
\end{equation}
Considering the long-term dynamics of the solution, eq. \eqref{appendix_eq:renewal_model_3_complex_3} can be rewritten as
\begin{align} \label{appendix_eq:renewal_model_3_complex_4}
E(t, 0) = \int_0^t b_{ME}\left(t, a_M\right) \int_0^{t - a_M} b_{IM}\left(t-a_M, a_I\right) I\left(t-a_M - a_I, 0\right) d a_I e^{-\mu a_M} d a_M 
\end{align}

Using \eqref{appendix_eq:boundary_on_infectious_model_three}, we have:
\begin{equation} \label{appendix_eq:renewal_model_3_complex_final}
\begin{aligned}
E(t, 0) &= \int_0^t b_{ME}\left(t, a_M\right) e^{-\mu a_M}  \int_0^{t - a_M} b_{IM}\left(t-a_M, a_I\right) \\ &\cdot  \int_0^{\infty} b_{E I}\left(t-a_M - a_I, a_E\right) E\left(t-a_M - a_I, a_E\right) d a_E d a_I d a_M 
\end{aligned}
\end{equation}
Now, in a similar approach to that taken in Section \ref{sec:infection_structured_model_2}, we derive a more conventional form of the renewal equation for the system.  The characteristic of the exposed human sub-population yields
\begin{align} \label{appendix_eq:characteristics_cases_exposed_model_three}
    E(t,a_E) &=
    \begin{cases}
    E(t - a_E,0) e^{-\int_{0}^{a_E} b_{EI}(t,s) ds}, & \text{if $a_E\leq t$}\\
        f_E(a_E - t) e^{-\int_{a_E - t}^{a_E} b_{EI}(t,s) ds}, & \text{if  $a_E > t$}
  \end{cases} 
  \end{align}
Focusing on long-term dynamics (i.e. $t \gg a_E$ and hence $f_E(a_E - t) e^{-\int_{a_E - t}^{a_E} b_{EI}(t,s) ds} = 0$), we have the following renewal equation (where the incidence of new infections is a function of the new infections at earlier times):
\small
\begin{equation}\label{appendix_eq:renewal_model_3_simpler_3}
\begin{aligned}
E(t, 0) &= \int_0^t b_{ME}\left(t, a_M\right) e^{-\mu a_M} \int_0^{t - a_M} b_{IM}\left(t-a_M, a_I\right) \\&  \cdot \int_0^{t - a_M - a_I} b_{E I}\left(t-a_M - a_I, a_E\right)  E\left(t-a_M - a_I - a_E, 0\right) e^{-\int_{0}^{a_E} b_{EI}(t - a_M - a_I,s) ds} d a_E d a_I  d a_M 
\end{aligned}
\end{equation}

\normalsize

As we are primarily considered with a long-term solution of our system, similar to Section \eqref{sec:infection_structured_model_2}, we derive a solution for the exposed human compartment of the system of the form: 
\begin{equation}\label{{appendix:separable_soln_exposed_model_three}}
    E(t,a_E) = p(a_E) e^{\gamma t}.
\end{equation}
This solution states that the age distribution of exposed individuals is altered by a factor which grows or decays with time depending on whether $\gamma > 0$ or $\gamma < 0$ respectively. 
So, if we are to assume such a separable solution (to derive analytical results), then we need that $b_{E I}\left(t, a_{E}\right) = b_{E I}\left(a_{E}\right) \forall t$. So, we assume that the (per capita) rate at which an exposed individual becomes infectious is independent of calendar time and depends only on time since exposure.

We first substitute the expression from eq.\eqref{appendix:separable_soln_exposed_model_three} into eq. \eqref{appendix_eq:exposed_model_three} to obtain:
\begin{equation}
    \gamma p(a_E) e^{\gamma t} + \frac{d p(a_E)}{da_E} e^{\gamma t} = - b_{E I}\left(a_{E}\right) p(a_E) e^{\gamma t} ,
\end{equation}
which implies
\begin{equation}\label{appendix_eq:model_3_age}
    p(a_E) = p(0) e^{-\gamma \left(a_E + \int_0^{a_E} b_{EI}(t, s) ds\right)}.
\end{equation}
Substituting the long-term solution and eq. \eqref{appendix_eq:model_3_age} into eq. \eqref{appendix_eq:renewal_model_3_simpler_3} yields:
\small
\begin{equation}
\begin{aligned}
    p(0) e^{\gamma t} &=    \int_0^t b_{ME}\left(t, a_M\right) e^{-\mu a_M}  \int_0^{t - a_M} b_{IM}\left(t-a_M, a_I\right) \\ &\cdot  \int_0^{t - a_M - a_I} b_{E I}\left(a_E\right) e^{-\int_{0}^{a_E} b_{EI}(s) ds} p(0) e^{\gamma (t-a_M-a_I-a_E)} d a_E d a_I d a_M 
\end{aligned}
\end{equation}
\normalsize
After cancelling $p(0) e^{\gamma t}$ from both sides of the equation, we have:
\begin{equation}
\begin{aligned}
    1 &= \int_0^t b_{ME}\left(t, a_M\right) e^{-\mu a_M}  \int_0^{t - a_M} b_{IM}\left(t-a_M, a_I\right) \\ &\cdot  \int_0^{t - a_M - a_I} b_{E I}\left(a_E\right) e^{-\int_{0}^{a_E} b_{EI}(s) ds} e^{-\gamma (a_M+a_I + a_E)} d a_E d a_I d a_M =: \phi(\gamma)
\end{aligned}
\end{equation}
Now, as $\phi(\gamma)$ is again a monotonically decreasing function of $\gamma$, identical to the system in Section \ref{sec:infection_structured_model_2}, we consider $\phi(0)$ to derive a reproduction number formula:
\begin{equation}
\begin{aligned}
\label{appendix_eq:reproduction_number_model_3_simple}
    R(t) &=  \int_0^t b_{ME}\left(t, a_M\right) e^{-\mu a_M}  \int_0^{t - a_M} b_{IM}\left(t-a_M, a_I\right) \\ &\cdot  \int_0^{t-a_M - a_I} b_{E I}\left(a_E\right) e^{-\int_{0}^{a_E} b_{EI}(s) ds} d a_E d a_I d a_M
\end{aligned}
\end{equation}
%


Let $
R_{E I}(t) :=\int_0^t b_{E I}\left(a_E\right) e^{-\int_0^{a_E} b_{E I}(s) d s} d a_E
$ which again can be used to derive a normalised version of the equation: 
$1 = \int_{0}^{t} \beta_{EI}(a_E) d a_E$ where $\beta_{EI}(a_E) := b_{E I}\left(a_E\right) e^{-\int_0^{a_E} b_{E I}(s) d s}/R_{EI}(t)$. 

Replicating a similar procedure for the transitions between the mosquito and exposed human sub-populations, we define $R_{ME}(t):=\int_{0}^{t} b_{ME}(t,a_M)  e^{-\mu a_M} d a_M$, which leads to the  equation: $1 = \int_{0}^{t} \beta_{ME}(t,a_M) d a_M$ where $\beta_{ME}(t,a_M) := b_{ME}(t,a_M) e^{-\mu a_M}/R_{ME}(t)$. 

We can also write $R_{IM}(t):=\int_{0}^{t} b_{IM}(t,a_I) d a_I$, which leads to the  equation: $1 = \int_{0}^{t} \beta_{IM}(t,a_I) d a_I$ where $\beta_{IM}(t,a_I) := b_{IM}(t,a_I)/R_{IM}(t)$.

\newpage

This suggests writing eq. \eqref{appendix_eq:reproduction_number_model_3_simple} as:
\begin{equation}
\begin{aligned}
R(t) & = R_{M E}(t) \mathbb{E}_{M E}{\left[R_{IM}(t - a_M) \int_0^{t-a_M} \beta_{IM }\left(t - a_M, a_I\right) R_{E I}\left(t-a_M - a_I\right) d a_I\right] } \\
& = R_{M E}(t) \mathbb{E}_{M E} \left[R_{IM}(t - a_M) \mathbb{E}_{I M} \left[R_{E I}\left(t-a_M - a_I\right)\right] \right] 
\end{aligned}
\end{equation}
where the expectations are defined as $\mathbb{E}_{M E}[X(t-a_M)]:=\int_{0}^t X(t-a_M) \beta_{ME}(t,a_M) da_M$ and 
$\mathbb{E}_{IM}[X(t-a_M-a_I)]:=\int_{0}^{t-a_M} X(t-a_M - a_I) \beta_{IM}(t-a_M,a_I) da_I$.


Alternatively, we have 
\begin{equation}
\begin{aligned}
R(t) &=  \int_0^t \int_0^{t - a_M} \int_0^{t - a_M - a_I} 
 b_{ME} \left(t, a_M\right) e^{-\mu a_M}  b_{IM}\left(t-a_M, a_I\right)  b_{E I}\left(a_E\right) \\ & \cdot e^{-\int_{0}^{a_E} b_{EI}(s) ds} d a_E d a_I d a_M \\
& = \int_0^t \int_0^{t - a_M} \int_0^{t - a_M - a_I}  K(t,a_E,a_I,a_M) d a_E d a_I  d a_M
\end{aligned} 
\end{equation}
\normalsize
which suggests definition of a probability density for the generation time at time $t$, \begin{equation}
    g(t,a_E, a_I,a_M):=K(t,a_E,a_I,a_M)/R(t).
\end{equation} 

We can write eq. \eqref{appendix_eq:renewal_model_3_simpler_3} as:
\begin{equation}\label{appendix_eq:renewal_2d_model_3_final}
\begin{aligned}
     E(t,0) &= R(t)\int_{0}^{t} \int_{0}^{t-a_M} \int_{0}^{t-a_M-a_I} g(t,a_E, a_I,a_M) E\left(t-a_M - a_I - a_E, 0\right) d a_E d a_I d a_M
\end{aligned}
\end{equation}
%

%



\newpage
\subsection{Relation to the simple ubiquitous renewal equation frameworks} \label{sec:relation_model_3_to_typical_renewal}
Similar to Section \ref{sec:infection_structured_model_2}, we can relate our derived renewal equation is to the renewal equations frequently used in directly transmitted infectious diseases. 

Let $\tau = a_E + a_I + a_M$ denote the total delay from primary human infection to secondary transmission, derived by addition of the times-between-states (i.e. the stages of the transmission cycle). 

Focusing on the integral 
\begin{equation*}
    \int_{0}^{t} \int_{0}^{t-a_M} \int_{0}^{t-a_M-a_I} g(t,a_E, a_I,a_M) E\left(t-a_M - a_I - a_E, 0\right) d a_E d a_I d a_M
\end{equation*}
from eq. \eqref{appendix_eq:renewal_2d_model_3_final}, we will now use a change of variables to convert $(a_E, a_I, a_M) \rightarrow (a_E, a_I, \tau)$. Using bijective functions $l(a_E, a_I, \tau) = a_E$, $m(a_E, a_I, \tau) = a_I$ and $n(a_E, a_I, \tau) = a_M = \tau - a_E - a_I$, we have:
\begin{equation}
    \begin{aligned}
   \int_{0}^{t} \int_{0}^{t-a_M} \int_{0}^{t-a_M-a_I} g(t,a_E, a_I,a_M) E\left(t-a_M - a_I - a_E, 0\right) d a_E d a_I d a_M \\
    =    \int_{0}^{t} \int_{0}^{\tau} \int_{0}^{\tau - a_I} g(t,l(a_E, a_I, \tau, m(a_E, a_I, \tau),n(a_E, a_I, \tau)) \\ \cdot  E(t-n(a_E, a_I, \tau)- m(a_E, a_I, \tau) - l(a_E, a_I, \tau),0) \left|\frac{\partial(a_E, a_I, a_M)}{\partial(a_E, a_I, \tau)}\right|d a_E d a_I d\tau \\
    = \int_{0}^{t} \int_{0}^{\tau} \int_{0}^{\tau - a_I} g(t,a_E, a_I, \tau-a_E-a_I)  E(t-a_E-a_I-(\tau-a_E-a_I),0)d a_E d a_I d\tau
    \\ = \int_{0}^{t} E(t-\tau,0) \int_{0}^{\tau} \int_{0}^{\tau - a_I} g(t,a_E, a_I, \tau-a_E-a_I)  d a_E d a_I d\tau \\
      = \int_{0}^{t} E(t-\tau,0) \int_{0}^{\tau} [\tilde{g}(t,a_E, a_I, \tau-a_E-a_I)]_0^{\tau - a_I} da_I  d\tau \\
                        = \int_{0}^{t} E(t-\tau,0) w(t,\tau)\ d\tau
\end{aligned}
\end{equation}

where $ \left|\frac{\partial(a_E, a_I, a_M)}{\partial(a_E, a_I,\tau)}\right| = 1$ denotes the Jacobian (i.e. the determinant of the Jacobian matrix), $\tilde{g}(t,a_E, a_I, \tau-a_E-a_I)$ is the antiderivative of $g(t,a_E, a_I, \tau-a_E-a_I)$ with respect to $a_E$, and $w(t, \tau)$ is a definite integral (of a definite integral of $\tilde{g}$ with respect to $a_I$ ) defined as  $w(t, \tau):= \int_{0}^{\tau} [\tilde{g}(t,a_E, a_I, \tau-a_E-a_I)]_0^{\tau - a_I} d a_I$. Here, $[\tilde{g}(t,a_E, a_I, \tau-a_E-a_I)]_0^{\tau - a_I}$ is a function of $(t, a_I, \tau)$ as the dependence on $a_E$ has been integrated (i.e. marginalised) out from the expression.


Similar to eq. \eqref{appendix_eq:renewal_2d_model_2_additive_delay_inhomogeneous_gi}, we can relate eq. \eqref{appendix_eq:renewal_2d_model_3_final} and hence our framework to the ubiquitous renewal equation framework of directly transmitted infectious diseases, albeit with a time-varying generation time $w(t, \tau)$: 

\begin{equation}\label{appendix_eq:renewal_2d_model_3_additive_delay_inhomogeneous_gi}
     E(t,0) = R(t)\int_{0}^{t} E(t-\tau,0) w(t, \tau) d\tau.
\end{equation}

In short, the relation (and compression to a 1-D marginal density $w(t, \tau)$ at time $t$), has been established by i) the Jacobian-based change of variable (to frame our integral equation in terms of the sum of the times-between states) and ii) $C-1=2$ (inner) integration steps (where $C=3$ is the number of model compartments) to marginalise out the first two times-between state variables.

Once again, assuming a constant density for the generation time over calendar time $g(t,a_E, a_I,a_M) = g(a_E, a_I,a_M) \forall t$, we arrive at the ubiquitous renewal equation framework of directly transmitted infectious diseases:

\begin{equation}\label{appendix_eq:renewal_2d_model_3_additive_delay_homogeneous_gi}
          E(t,0) = R(t)\int_{0}^{t} E(t-\tau,0) w(\tau) d\tau.
\end{equation}

Here, assuming a constant generation time over calendar time is equivalent to assuming that the relative contribution of each times-between-state triplet ($a_E,a_I,a_M$) to secondary human infections is fixed throughout the course of an outbreak. In further detail, this means that the numerator (the instantaneous kernel at time $t$) and denominator (the reproduction number at time $t$) of the following expression always appear in fixed proportions:
\begin{equation}
\begin{aligned}
    g(a_E, a_I,a_M) &= \frac{K(t,a_E,a_I,a_M)}{R(t)} \\  &= \frac{b_{ME} \left(t, a_M\right) e^{-\mu a_M}  b_{IM}\left(t-a_M, a_I\right)  b_{E I}\left(a_E\right) e^{-\int_{0}^{a_E} b_{EI}(s) ds}}{\int_{0}^{t} \int_{0}^{t-a_M} \int_{0}^{t-a_M-a_I}  K(t,a_E,a_I,a_M)  d a_E d a_I d a_M}. 
\end{aligned}
\end{equation}

Again, as we have described in detail in Section \ref{sec:infection_structured_model_2}, the assumption of calendar-time-homogeneous relative contributions of times-in-state triplet ($a_E,a_I,a_M$) may be unrealistic for transmission of vector-borne diseases. Nevertheless, via eq. \eqref{appendix_eq:renewal_2d_model_3_additive_delay_homogeneous_gi}, we have formalised a relationship between our derived framework and the more conventional renewal equation frameworks.
\newpage

\section{Exposure-and-infection-age-structured populations of human and mosquitoes}\label{sec:exposed_and_infection_structured_model_4_appendix}
\subsection{Modelling framework}
We now provide further details of the framework described in Section \ref{sec:exposed_and_infection_structured_model_4}. Recall that the system was described with the following linear PDEs for the exposed ($E$) and infectious ($I$) human populations, and exposed ($W$) and infectious ($V$) mosquito populations
\begin{align}
\frac{\partial E}{\partial t}+\frac{\partial E}{\partial a_E} &=-b_{E I}\left(t, a_{E}\right) E\left(t, a_{E}\right) \label{appendix_eq:exposed_model_four}\\
\frac{\partial I}{\partial t}+\frac{\partial I}{\partial a_I}&=0 
\label{appendix_eq:infectious_model_four}
\\
\frac{\partial W}{\partial t}+\frac{\partial W}{\partial a_W}&=-b_{W V}\left(t, a_{W}\right) W\left(t, a_{W}\right) -\mu W\left(t, a_{W}\right)
\label{appendix_eq:mosquito_exposed_model_four} 
\\
\frac{\partial V}{\partial t}+\frac{\partial V}{\partial a_V}&=-\mu V\left(t, a_{V}\right) \label{appendix_eq:mosquito_infectious_model_four}
\end{align}
where again each $b_{AB}(t,a_A) \geq 0$ is a time- and age-dependent birth function that represents the rate at which new individuals in sub-population $B$ are generated at calendar time $t$ by an individual of age $a_A$ in sub-population $A$. $\mu \geq 0$ is the mosquito death rate for both exposed and infectious mosquitoes, which we assume is age-independent \citep{lambert_meta-analysis_2022}.

We close the system with the following initial and boundary conditions
\begin{align}
E(t, 0) &=\int_0^\infty b_{VE}\left(t, a_V\right) V\left(t, a_V\right) d a_V  \label{appendix_eq:boundary_on_exposed_model_four}\\
E\left(0, a_E\right) &= f_E\left(a_E\right) \\
I(t, 0) &=\int_0^{\infty} b_{E I}\left(t, a_E\right) E\left(t, a_E\right) d a_E \label{appendix_eq:boundary_on_infectious_model_four} \\
I\left(0, a_I\right) &=f_I\left(a_I\right) \\
W(t, 0) &=\int_0^{\infty} b_{IW}\left(t, a_I\right) I\left(t, a_I\right) d a_I \label{appendix_eq:boundary_on_exposed_mosquitoes_model_four} \\
W\left(0, a_W\right) &=f_W\left(a_W\right)  \\
V(t, 0) &=\int_0^{\infty} b_{WV}\left(t, a_W\right) W\left(t, a_W\right) d a_W \label{appendix_eq:boundary_on_infectious_mosquitoes_model_four} \\
V\left(0, a_V\right) &=f_V\left(a_V\right)  \label{appendix_eq:initial_on_infectious_mosquitoes_model_four}
\end{align}
where $f_E(a_E), f_I(a_I), f_W(a_W),$ and $f_V(a_V)$ are the initial densities for the exposed human, infectious human, exposed mosquito, and infectious mosquito populations.

For this system, from the characteristics for the following sub-populations, we have:
\begin{align} \label{appendix_eq:characteristics_cases_exposed_model_four}
    E(t,a_E) &=
    \begin{cases}
    E(t - a_E,0) e^{-\int_{0}^{a_E} b_{EI}(t,s) ds}, & \text{if $a_E\leq t$}\\
        f_E(a_E - t) e^{-\int_{a_E - t}^{a_E} b_{EI}(t,s) ds}, & \text{if  $a_E > t$}
  \end{cases} \\
  I(t,a_I) &=
    \begin{cases}
    I(t - a_I,0) , & \text{if $a_I\leq t$}\\
f_I(a_I - t), & \text{if  $a_I > t$}
  \end{cases} \label{appendix_eq:characteristics_cases_infectious_model_four}\\
W\left(t, a_W\right) &= \begin{cases}W\left(t-a_W, 0\right) e^{-\int_0^{a_W} b_{WV}(t, r) d r - \mu a_W}, & \text{if $a_W\leq t$} \\
f_W\left(a_W-t\right) e^{-\int_{a_{w-t}}^{a_w} b_{WV}(t, r) d r - \mu t} , & \text{if  $a_W > t$}\end{cases} \label{appendix_eq:characteristic_equations_exposed_mosquitoes}\\
& V\left(t, a_V\right)= \begin{cases}V\left(t-a_V, 0\right) e^{-\mu a_V}, & \text{if $a_V\leq t$} \\
f_V\left(a_V-t\right) e^{-\mu t} , & \text{if  $a_V > t$}\end{cases}  \label{appendix_eq:characteristic_equations_infectious_mosquitoes}
&
\end{align}
\subsection{Deriving a renewal equation and time-varying reproduction number} 
We now provide further details of the derivation for our instantaneous reproduction number, eq. \eqref{eq:reprod_number_model_four_in_terms_of_density_in_manuscript}.
Once again, we model the length of time between primary human infection and secondary human transmission i.e. the interval between a newly exposed individual and the secondary exposed human originating from this infection. Rewriting eq. \eqref{appendix_eq:boundary_on_exposed_model_four}, we have:
\begin{equation}\label{appendix_eq:model_four_renewal_equation_1}
\begin{aligned}
E(t, 0) &=\int_0^t b_{VE}\left(t, a_V\right) V\left(t-a_V, 0\right) e^{-\mu a_V} d a_V \\ &+ \int_t^\infty b_{VE}\left(t, a_V\right) f_V\left(a_V-t\right) e^{-\mu t} d a_V
\end{aligned}
\end{equation}
Once again, we focus primarily on the long-time behaviour of the system, i.e. when $t\gg a_V$ and $f_V\left(a_V-t\right) e^{-\mu t}\approx 0$. We can ignore the second part of the equation, and also use eq. \eqref{appendix_eq:boundary_on_infectious_model_four} to write:
\begin{align}\label{appendix_eq:model_four_renewal_equation_2}
E(t, 0) &=\int_0^t b_{VE}\left(t, a_V\right) e^{-\mu a_V} \int_0^{\infty} b_{WV}\left(t-a_V, a_W\right) W\left(t-a_V, a_W\right)   d a_W d a_V 
\end{align}
Using the characteristics for the exposed mosquito population (eq. \eqref{appendix_eq:characteristic_equations_exposed_mosquitoes}), and again focusing on $t \gg a_W$ (such that $f_W\left(a_W-t\right) e^{-\int_{a_{w-t}}^{a_w} b_{WV}(t, r) d r - \mu t} \approx 0$) we have:
\begin{equation}\label{appendix_eq:model_four_renewal_equation_3}
\begin{aligned}
E(t, 0) &=\int_0^t b_{VE}\left(t, a_V\right) e^{-\mu a_V} \int_0^{t-a_V} b_{WV}\left(t-a_V, a_W\right) W\left(t-a_V - a_W, 0\right) \\ &\cdot e^{-\int_0^{a_W} b_{WV}(t-a_V, r) d r - \mu a_W} d a_W d a_V 
\end{aligned}
\end{equation}
Substituting in eq. \eqref{appendix_eq:boundary_on_exposed_mosquitoes_model_four}, we have:
\begin{equation}
\label{appendix_eq:model_four_renewal_equation_3.5}
\begin{aligned}
E(t, 0) &=\int_0^t b_{VE}\left(t, a_V\right) e^{-\mu a_V} \int_0^{t-a_V} b_{WV}\left(t-a_V, a_W\right)  \int_0^{\infty} b_{IW}\left(t-a_V - a_W, a_I\right) \\ &\cdot I\left(t-a_V - a_W, a_I\right)e^{-\int_0^{a_W} b_{WV}(t-a_V, r) d r - \mu a_W}  d a_I   d a_W d a_V 
\end{aligned}
\end{equation}
Now, using the characteristics for the infectious human population, and focusing on long-term dynamics, we have:
\begin{equation}
    \begin{aligned}\label{appendix_eq:model_four_renewal_equation_4}
E(t, 0) &=\int_0^t b_{VE}\left(t, a_V\right) e^{-\mu a_V} \int_0^{t-a_V} b_{WV}\left(t-a_V, a_W\right) e^{-\int_0^{a_W} b_{WV}(t-a_V, r) d r - \mu a_W} \\ &\cdot \int_0^{t-a_V-a_W} b_{IW}\left(t-a_V - a_W, a_I\right)   I\left(t-a_V - a_W - a_I, 0\right)     d a_I d a_W   d a_V 
\end{aligned}
\end{equation}

Using the boundary condition for the infectious human population, we have:
\begin{equation}\label{appendix_eq:model_four_renewal_equation_5}
    \begin{aligned}
E(t, 0) &=\int_0^t b_{VE}\left(t, a_V\right) e^{-\mu a_V} \int_0^{t-a_V} b_{WV}\left(t-a_V, a_W\right)  \int_0^{t-a_V-a_W} b_{IW}\left(t-a_V - a_W, a_I\right) \\ &\cdot e^{-\int_0^{a_W} b_{WV}(t-a_V, r) d r - \mu a_W} \int_0^{\infty} b_{E I}\left(t-a_V - a_W - a_I, a_E\right)  E\left(t-a_V - a_W - a_I, a_E\right) \\ & \cdot d a_E  d a_I  d a_W   d a_V
\end{aligned}
\end{equation}

\normalsize
We can rearrange eq. \eqref{appendix_eq:model_four_renewal_equation_5} to emphasise the role of the birth processes and relatedly, the ages of the source individuals at the times at which secondary births occur:
\begin{equation}\label{appendix_eq:model_four_renewal_equation_on_infectious_v1} 
\begin{aligned}
E(t, 0) &=\int_0^t \int_0^{t-a_V}  \int_0^{t-a_V-a_W}  \int_0^{\infty} b_{VE}\left(t, a_V\right)   b_{WV}\left(t-a_V, a_W\right) b_{IW}\left(t-a_V - a_W, a_I\right)  \\ &\cdot b_{E I}\left(t-a_V - a_W - a_I, a_E\right) e^{-\mu a_V} e^{-\int_0^{a_W} b_{WV}(t-a_V, r) d r - \mu a_W}    E\left(t-a_V - a_W - a_I, a_E\right) \\ & \cdot d a_E  d a_I  d a_W   d a_V
\end{aligned}
\end{equation}
\normalsize
Using eq. \eqref{appendix_eq:characteristics_cases_exposed_model_four}, focusing on long-term dynamics ($t \gg a_E$), we have the following renewal equation:
\small
\begin{equation}\label{appendix_eq:model_four_renewal_equation_on_infectious_v1}
\begin{aligned}  E(t, 0) &=\int_0^t \int_0^{t-a_V}  \int_0^{t-a_V-a_W}  \int_0^{t-a_V - a_W - a_I} b_{VE}\left(t, a_V\right)   b_{WV}\left(t-a_V, a_W\right) b_{IW}\left(t-a_V - a_W, a_I\right)  \\ &\cdot b_{E I}\left(t-a_V - a_W - a_I, a_E\right) e^{-\mu a_V} e^{-\int_0^{a_W} b_{WV}(t-a_V, r) d r - \mu a_W}    \\ & \cdot E(t-a_V - a_W - a_I - a_E,0) e^{-\int_{0}^{a_E} b_{EI}(t-a_V - a_W - a_I,s) ds} \\ & \cdot d a_E  d a_I  d a_W   d a_V & 
\end{aligned}
\end{equation}
\normalsize

We are primarily considered with a long-term solution of our system, and now, similar to Section \eqref{sec:exposed_and_infection_structured_model_3}, we derive a solution for the infectious human compartment of the system of the form: 
\begin{align}\label{appendix:separable_soln_exposed_model_four}
    E(t,a_E) = p(a_E) e^{\gamma t}.
\end{align}
This separable solution amounts to assuming that the age-since-exposure distribution of the exposed human compartment, is altered by a time-dependent factor which either grows ($\gamma > 0$) or decays ($\gamma<0$). So, if we are to assume such a separable solution (to derive analytical results), then we need that $b_{E I}\left(t, a_{E}\right) = b_{E I}\left(a_{E}\right) \forall t$. So, we assume that the (per capita) rate at which an exposed individual becomes infectious is independent of calendar time and depends only on time since exposure.

We first substitute the expression from eq.\eqref{appendix:separable_soln_exposed_model_four} into eq. \eqref{appendix_eq:exposed_model_four}to obtain:
\begin{equation}
    \gamma p(a_E) e^{\gamma t} + \frac{d p(a_E)}{da_E} e^{\gamma t} = - b_{E I}\left(a_{E}\right) p(a_E) e^{\gamma t} ,
\end{equation}


%
which implies
\begin{equation}\label{appendix_eq:model_4_age}
    p(a_E) = p(0) e^{-\gamma \left(a_E + \int_0^{a_E} b_{EI}(s) ds\right)}.
\end{equation}

%
Substituting the separable solution  (eq. \eqref{appendix:separable_soln_exposed_model_four}) and eq. \eqref{appendix_eq:model_4_age} into eq. \eqref{appendix_eq:model_four_renewal_equation_on_infectious_v1} yields:
\small
\begin{equation}
    \begin{aligned}
p(0) e^{\gamma t} &=\int_0^t \int_0^{t-a_V}  \int_0^{t-a_V-a_W}  \int_0^{t-a_V - a_W - a_I} b_{VE}\left(t, a_V\right)   b_{WV}\left(t-a_V, a_W\right) b_{IW}\left(t-a_V - a_W, a_I\right)  \\ &\cdot b_{E I}\left(a_E\right) e^{-\mu a_V} e^{-\int_0^{a_W} b_{WV}(t-a_V, r) d r - \mu a_W} e^{-\int_{0}^{a_E} b_{EI}(s) ds} \\ & \cdot p(0) e^{\gamma (t - a_V - a_W - a_I - a_E)} d a_E  d a_I  d a_W   d a_V 
\end{aligned}
\end{equation}
\normalsize
where we have used the assumption that $b_{E I}\left(t, a_{E}\right) = b_{E I}\left(a_{E}\right) \forall t$.

Upon cancelling $p(0) e^{\gamma t}$ from both sides of the equation, we have:
\small
\begin{equation}
\begin{aligned}
1 &=\int_0^t \int_0^{t-a_V}  \int_0^{t-a_V-a_W}  \int_0^{t-a_V - a_W - a_I} b_{VE}\left(t, a_V\right)   b_{WV}\left(t-a_V, a_W\right) b_{IW}\left(t-a_V - a_W, a_I\right)  \\ &\cdot b_{E I}\left(a_E\right) e^{-\mu a_V} e^{-\int_0^{a_W} b_{WV}(t-a_V, r) d r - \mu a_W} e^{-\int_{0}^{a_E} b_{EI}(s) ds} \\ & \cdot e^{-\gamma (a_V + a_W + a_I + a_E)}  d a_E  d a_I  d a_W   d a_V =: \phi(\gamma)
\end{aligned}
\end{equation}
\normalsize
Now, as $\phi(\gamma)$ is a decreasing function of $\gamma$, similar to the system in Section \ref{sec:infection_structured_model_2}, we have that the exponential growth/decay of human infections is controlled by the growth/decay parameter $\gamma$ alongside times spent in each of the four individual compartments. 

We consider $\phi(0)$ to derive a reproduction number formula:
\small
\begin{equation}\label{appendix_eq:reproduction_number_model_4_simple}
\begin{aligned}
R(t) &=\int_0^t \int_0^{t-a_V}  \int_0^{t-a_V-a_W}  \int_0^{t-a_V - a_W - a_I} b_{VE}\left(t, a_V\right)   b_{WV}\left(t-a_V, a_W\right) b_{IW}\left(t-a_V - a_W, a_I\right)  \\ &\cdot b_{E I}\left(a_E\right) e^{-\mu a_V} e^{-\int_0^{a_W} b_{WV}(t-a_V, r) d r - \mu a_W}\\ & \cdot e^{-\int_{0}^{a_E} b_{EI}(t-a_V - a_W - a_I,s) ds} d a_E  d a_I  d a_W   d a_V
\end{aligned}
\end{equation}
\normalsize


Alternatively, we can write:
 \begin{equation}
     \label{appendix_eq:reprod_number_model_four_in_terms_of_density}
 \begin{aligned}
R(t) &= \int_0^t \int_0^{t-a_V}  \int_0^{t-a_V-a_W}  \int_0^{t-a_V - a_W - a_I} b_{VE}\left(t, a_V\right)   b_{WV}\left(t-a_V, a_W\right) b_{IW}\left(t-a_V - a_W, a_I\right)  \\ &\cdot b_{E I}\left(a_E\right) e^{-\mu a_V} e^{-\int_0^{a_W} b_{WV}(t-a_V, r) d r - \mu a_W} \\ & \cdot e^{-\int_{0}^{a_E} b_{EI}(s) ds} d a_E  d a_I  d a_W   d a_V \\ 
&= \int_0^t \int_0^{t-a_V}  \int_0^{t-a_V-a_W}  \int_0^{t-a_V - a_W - a_I}K(t,a_E, a_I,a_W,a_V) d a_E  d a_I  d a_W   d a_V
\end{aligned}
\end{equation}

which now suggests definition of a four-dimensional probability density $g(t,a_E, a_I,a_W,a_V):=K(t,a_E, a_I,a_W,a_V)/R(t)$. As our instantaneous reproduction number defines the expected number of secondary human infections produced by an individual initially infected at time $t$ (if conditions remained constant), $K$ denotes the number of secondary human infections are generated at calendar time $t$ for various individual times spent (represented as ages) in each state/compartment, whilst $g$ denotes the normalised kernel (otherwise known as the instantaneous generation time \citep{park_importance_2022}) which represents the relative contribution of past human infections, previously exposed humans, previously exposed mosquitoes, and currently infectious mosquitoes (across their various times within compartments i.e. ages) to the current number of humans currently infectious. In each case, for either $g, K,$ or $R$, we are describing a counterfactual scenario of contributions to the epidemic where the conditions would remain constant.

\newpage

\subsection{Between-stage reproduction numbers}\label{appendix:between_comp}
As before, focusing on generation events between stages, we define  
\begin{equation}
\begin{aligned}
R_{V E}(t) &:=\int_0^t b_{VE}\left(t, a_V\right)e^{-\mu a_V} d a_V,  \\
\beta_{VE}(a_V) &:= \frac{b_{VE}\left(t, a_V\right)e^{-\mu a_V}}{R_{VE}(t)},
\end{aligned}
\end{equation}
\normalsize
where $1 = \int_{0}^{t} \beta_{VE}(t,a_V) d a_V$. Replicating similar procedures for other sub-populations, we also define 
\begin{equation}
\begin{aligned}
R_{IW}(t) &:=\int_{0}^{t} b_{IW}(t,a_I)d a_I,  \\
\beta_{IW}(t,a_I) &:= \frac{b_{IW}(t,a_I)}{R_{IW}(t)},
\end{aligned}
\end{equation}
\normalsize
where $1 = \int_{0}^{t} \beta_{IW}(t,a_I) d a_I$.  Also, let  
\begin{equation}
\begin{aligned}
R_{WV}(t) &:= \int_{0}^{t} b_{WV}(t,a_W) e^{-\int_0^{a_W} b_{WV}\left(t, r\right) d r}d a_W, \\
\beta_{WV}(t,a_W) &:= \frac{b_{WV}(t,a_W) e^{-\int_0^{a_W} b_{WV}(t, r) d r - \mu a_W}}{R_{WV}(t)},
\end{aligned}
\end{equation}
\normalsize
where $1 = \int_{0}^{t} \beta_{WV}(t,a_W) d a_W$. Finally, let 
\begin{equation}
    \begin{aligned}
        R_{EI}(t) &:= {\int_{0}^{t} b_{E I}\left(a_E\right) e^{-\int_0^{a_E} b_{E I}(a_E) d s}da_E} \\
        \beta_{EI}(t,a_E) &:= \beta_{EI}(a_E) = \frac{b_{E I}\left(a_E\right) e^{-\int_0^{a_E} b_{E I}(a_E) d s}}{R_{EI}(t)}
    \end{aligned}
\end{equation}
where $1 = \int_{0}^{t} \beta_{EI}(t,a_E) d a_E$.
So, we have

\small
\begin{equation}
\begin{aligned}
\beta_{EI}(t,a_E) &= \beta_{EI}(a_E)  = \frac{K_{EI}(a_E)}{R_{EI}(t)}, \\
\beta_{IW}(t, a_I) &= \frac{b_{IW}(t, a_I)}{R_{IW}(t)} = \frac{K_{IW}(t, a_I)}{R_{IW}(t)}, \\
\beta_{WV}(t,a_W) &= \frac{b_{WV}(t,a_W) e^{-\int_0^{a_W} b_{WV}\left(t, r\right) d r - \mu a_W}}{R_{WV}(t)} = \frac{K_{WV}(t, a_W)}{R_{WV}(t)}, \\
\beta_{VE}(t,a_V) &= \frac{b_{V E}\left(t, a_V\right)  e^{-\mu a_V}}{R_{VE}(t)} = \frac{K_{VE}(t, a_V)}{R_{VE}(t)}.
\end{aligned}
\end{equation}
\normalsize
Each  $\beta_{CD}(t, a_C)$ defines a probability density (integrating to 1 at each calendar time $t$) for the generation time between stages ($C$ and $D$). It is obtained by normalising the corresponding instantaneous kernels $K_{CD}(t, a_C)$. $\beta_{CD}(t, a_C)$ determines the distribution (at calendar time $t$) for the times taken from initial entry (into compartment $C$) to the generation of a new individual in the other compartment ($D$). The time-varying nature means that the relative contributions of different times-in-state ($a_C$) vary at different calendar times. This may happen in our setting of vector-borne diseases (e.g. under time variations caused by climatic conditions or changing totals of susceptible and infectious hosts and vectors). For example, we capture time variance in the extrinsic incubation period over calendar time via $\beta_{WV}(t,a_W)$, measuring from initial mosquito exposure ($W$) to infectious mosquito state ($V$) the relative contribution of each age $a_W$ (time-since-mosquito-exposure) to the generation of a newly infectious mosquito.

So, we can rewrite eq. \eqref{eq:reprod_number_model_four_in_terms_of_density_in_manuscript} as
\small
\begin{equation*}
    \begin{aligned}
R(t) &=  R_{VE}(t) \int_0^t \beta_{VE}(t, a_V) R_{WV}(t-a_V) \int_0^{t-a_V}  \beta_{WV}\left(t-a_V, a_W\right) R_{IW}\left(t-a_V - a_W\right) \\ &\cdot \int_0^{t-a_V-a_W}  \beta_{IW}\left(t-a_V - a_W, a_I\right) R_{EI}(t-a_V - a_W - a_I)  d a_I  d a_W   d a_V 
\end{aligned}
\end{equation*}
\normalsize

and using expectations defined as 
\begin{equation*}
    \begin{aligned}
\mathbb{E}_{VE}[X(t - a_V)] &:=\int_{0}^t X(t-a_V) \beta_{VE}(t,a_V) da_V, \\ 
\mathbb{E}_{WV}[X(t-a_V-a_W)] &:=\int_{0}^{t-a_V} X(t-a_V-a_W) \beta_{WV}(t-a_V,a_W) da_W, \\
\mathbb{E}_{IW}[X(t-a_V-a_W-a_I)] &:=\int_{0}^{t-a_V-a_W} X(t-a_V-a_W -a_I) \beta_{IW}(t-a_V-a_W,a_I) da_I, \\ 
\end{aligned}
\end{equation*}

we also have:
\begin{align}
R(t) &=R_{V E}(t)\mathbb{E}_{V E}\left[R_{WV}(t-a_V) \mathbb{E}_{W V}\left[ R_{IW}(t-a_V-a_W) \mathbb{E}_{I W} \left[R_{E I}\left(t-a_V-a_W-a_I\right)  \right] \right]\right]. 
\end{align}
Again, the instantaneous reproduction number is an expectation of between-stage reproduction numbers across stages of the transmission cycle (the cycle required between two successive human infections). The between-stage reproduction numbers are evaluated with upper limits determined by calendar times at which the subsequent stage of the transmission cycle begins. The expectations are taken with respect to the generation time between stages  of the next stage of the transmission cycle. The generation times are evaluated using ages of the sub-populations at the calendar times at which onward generation events in the transmission cycle occur (see Figure \ref{fig:figure_1}d: $a_V$ at time $t$, $a_W$ at time $t-a_V$, and $a_I$ at time $t-a_V-a_W$).

\subsection{Application to simulated and real-world data}\label{appendix:application}
We provide here further details of the Monte Carlo sampling, distributional assumptions, and results from the application in Section \ref{sec:exposed_and_infection_structured_model_4}. The distributional assumptions were all based on existing work by \citet{siraj_temperature_2017}.

In \citet{siraj_temperature_2017}, the time from exposed (E) to infectious (I) was approximated by the intrinsic incubation period (the time from human infection to symptom onset). Based on results from a previous modelling review \citep{chan_incubation_2012}, the intrinsic incubation period was modelled using a log-normal distribution with mean of 5.97 days and standard deviation of 1.64 days. Denote this probability density as $\beta_{EI}(t, a_E)$.

For the time from infectious human (I) to infected mosquito (W), \citet{siraj_temperature_2017} used a normal probability density function multiplied by a calibration/scaling factor to fit to previously published experimental data \citep{nishiura_natural_2007}. Denote this probability density as $\beta_{IW}(t, a_I)$. 

Similar to the human population, the time from infected (W) to infectious (V) mosquito was approximated in \citet{siraj_temperature_2017} using the extrinsic incubation period (EIP), based on results from the same modelling review  \citep{chan_incubation_2012}. This EIP was modelled with a log-normal distribution with a shape parameter of 4.9 and scale parameter of $e^{2.9-0.08\cdot T}$ where $T$ is the observed temperature. Denote this probability density as $\beta_{WV}(t, a_W)$. 

The period from the infectious mosquito (V) to infected human (E) was modelled in \citet{siraj_temperature_2017} using an existing temperature- and age-dependent model \citep{brady_modelling_2013} with an additional rate of extrinsic mortality (of 0.089 per day). Denote this probability density as $\beta_{VE}(t, a_V)$. 

So, for the existing approach, at each iteration $i$ of our Monte Carlo sampling scheme, we can generate sample values $a_E^{(i)}, a_I^{(i)}, a_W^{(i)},$ and $a_V^{(i)}$ using the distributions described by $\beta_{EI}(t, a_E), \beta_{IW}(t, a_I), \beta_{WV}(t, a_W)$, and $ \beta_{VE}(t, a_V)$. For the stage-specific approach, at each iteration $i$ of our Monte Carlo sampling scheme, we generate sample values $a_V^{(i)}$ using $\beta_{VE}(t, a_V)$. Then, we sample $a_W^{(i)}$ using $\beta_{WV}(t-a_V^{(i)}, a_W)$,  $a_I^{(i)}$ using $\beta_{IW}(t-a_V^{(i)} - a_W^{(i)}, a_W)$, and $a_E^{(i)}$ using $\beta_{EI}(t-a_V^{(i)} - a_W^{(i)}-a_I^{(i)}, a_E)$. For either approach (stage-specific or existing), at iteration $i$, we let $\tau^{(i)} = a_E^{(i)} + a_I^{(i)} + a_W^{(i)} + a_V^{(i)}$. Repeating this procedure for $i \in (1, \ldots ,2000)$ produces 2000 Monte Carlo samples to approximate the two-dimensional generation time distribution $w(t, \tau)$.

\begin{figure}[t]
\centering
\includegraphics[scale = 0.175]{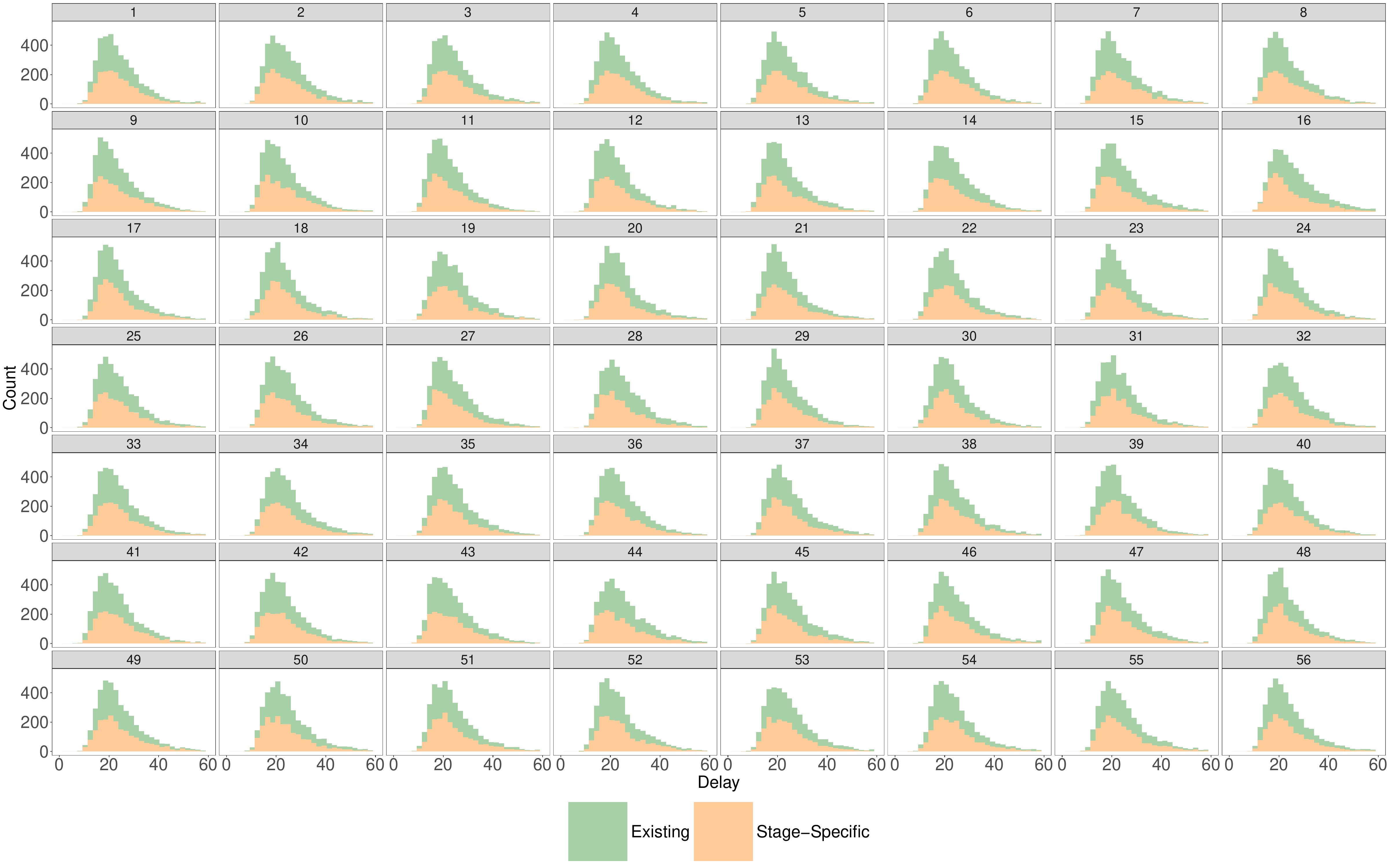}
\caption{\label{fig:amk_both_gi_histogram} \footnotesize \textbf{Time-varying generation times in Ang Mo Kio, Singapore}. For individual calendar times, represented by numbers in facet labels, we present histograms of Monte Carlo samples from the generation time distribution under different temperature-dependent assumptions. For the existing (light blue) approach, we allow the transmission cycle to change every day based on the temperature observed that day. Our stage-specific approach means that the generation time distribution can vary based on temperatures observed for calendar times when the estimated (by Monte Carlo samples) generation events occurred.}
\end{figure}
\begin{figure}[t]
\centering
\includegraphics[scale = 0.175]{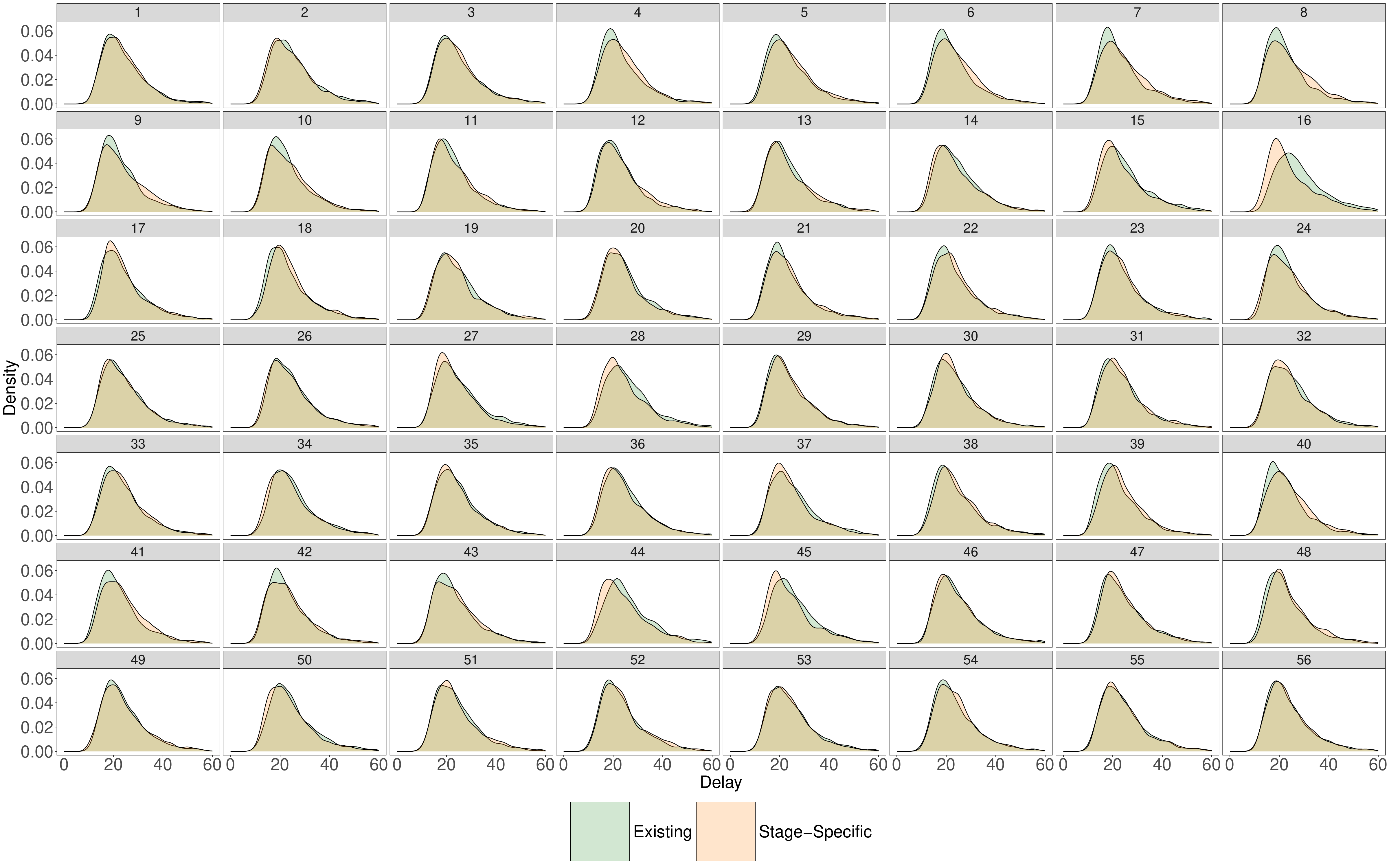}
\caption{\label{fig:amk_both_gi_density} \footnotesize \textbf{Time-varying generation times in Ang Mo Kio, Singapore}. For individual calendar times, represented by numbers in facet labels, we present kernel density estimates of Monte Carlo samples from the generation time distribution under different temperature-dependent assumptions. For the existing (light blue) approach, we allow the transmission cycle to change every day based on the temperature observed that day. Our stage-specific approach means that the generation time distribution can vary based on temperatures observed for calendar times when the estimated (by Monte Carlo samples) generation events occurred.}
\end{figure}
\begin{figure}[t]
\centering
\includegraphics[scale = 0.17]{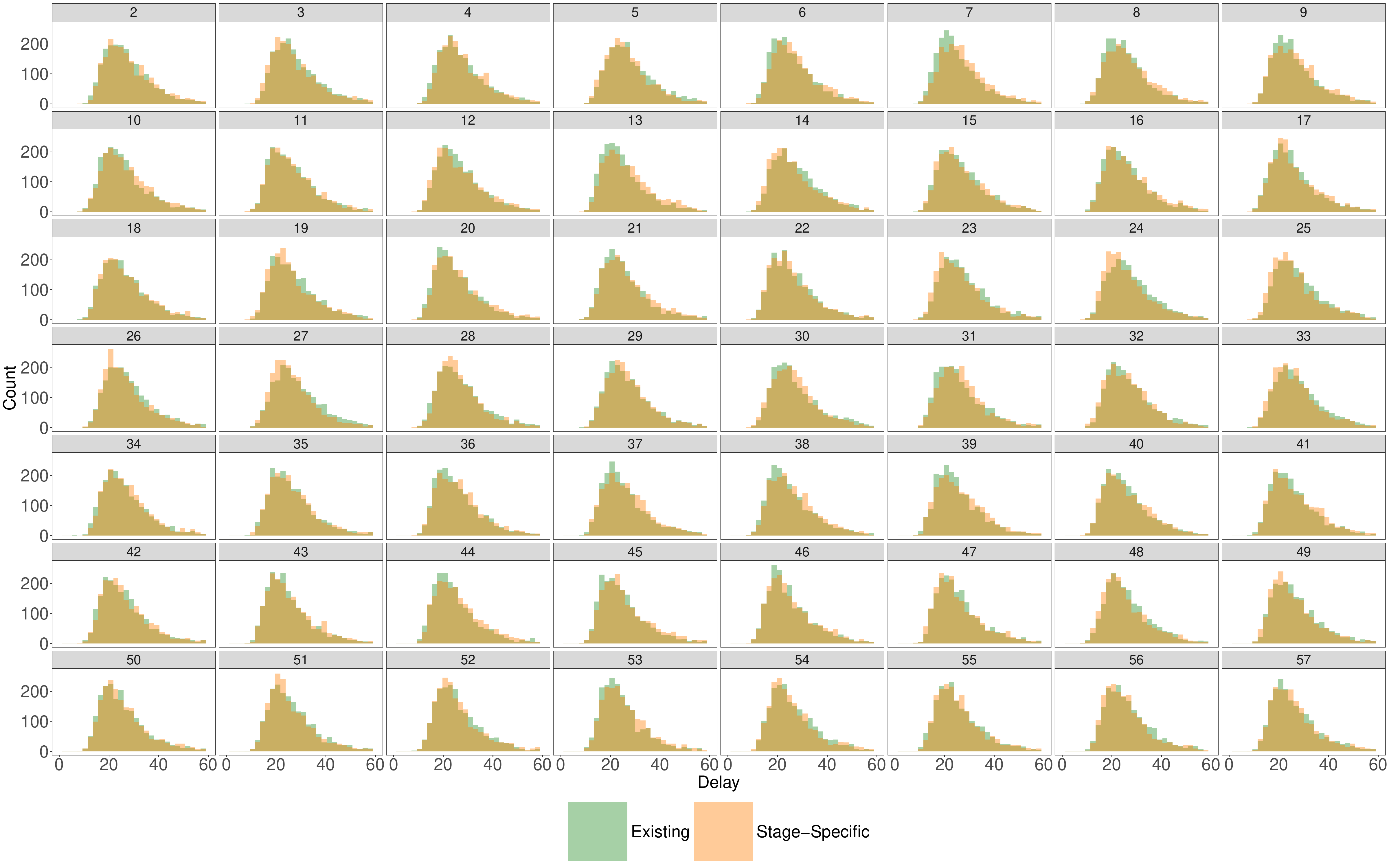}
\caption{\label{fig:sjt_both_gi_histogram} \footnotesize \textbf{Time-varying generation times in São João do Tauape, Fortaleza, Brazil}. For individual calendar times, represented by numbers in facet labels, we present histograms of Monte Carlo samples from the generation time distribution under different temperature-dependent assumptions. For the existing (light blue) approach, we allow the transmission cycle to change every day based on the temperature observed that day. Our stage-specific approach means that the generation time distribution can vary based on temperatures observed for calendar times when the estimated (by Monte Carlo samples) generation events occurred.}
\end{figure}
\begin{figure}[t]
\centering
\includegraphics[scale = 0.17]{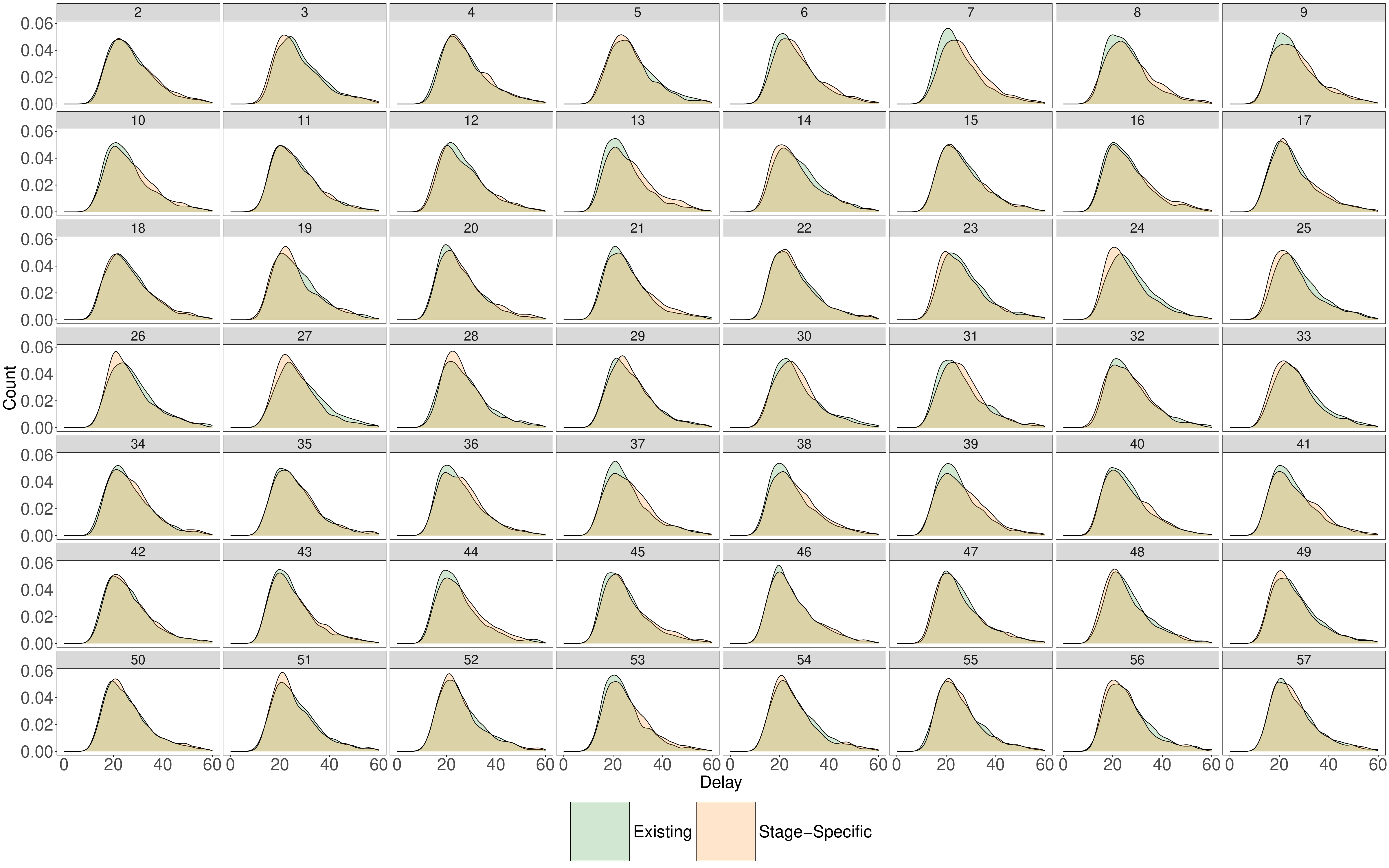}
\caption{\label{fig:sjt_both_gi_density} \footnotesize \textbf{Time-varying generation times in São João do Tauape, Fortaleza, Brazil}. For individual calendar times, represented by numbers in facet labels, we present kernel density estimates of Monte Carlo samples from the generation time distribution under different temperature-dependent assumptions. For the existing (light blue) approach, we allow the transmission cycle to change every day based on the temperature observed that day. Our stage-specific approach means that the generation time distribution can vary based on temperatures observed for calendar times when the estimated (by Monte Carlo samples) generation events occurred.}
\end{figure}
\begin{figure}[t]
\centering
\includegraphics[scale = 0.17]{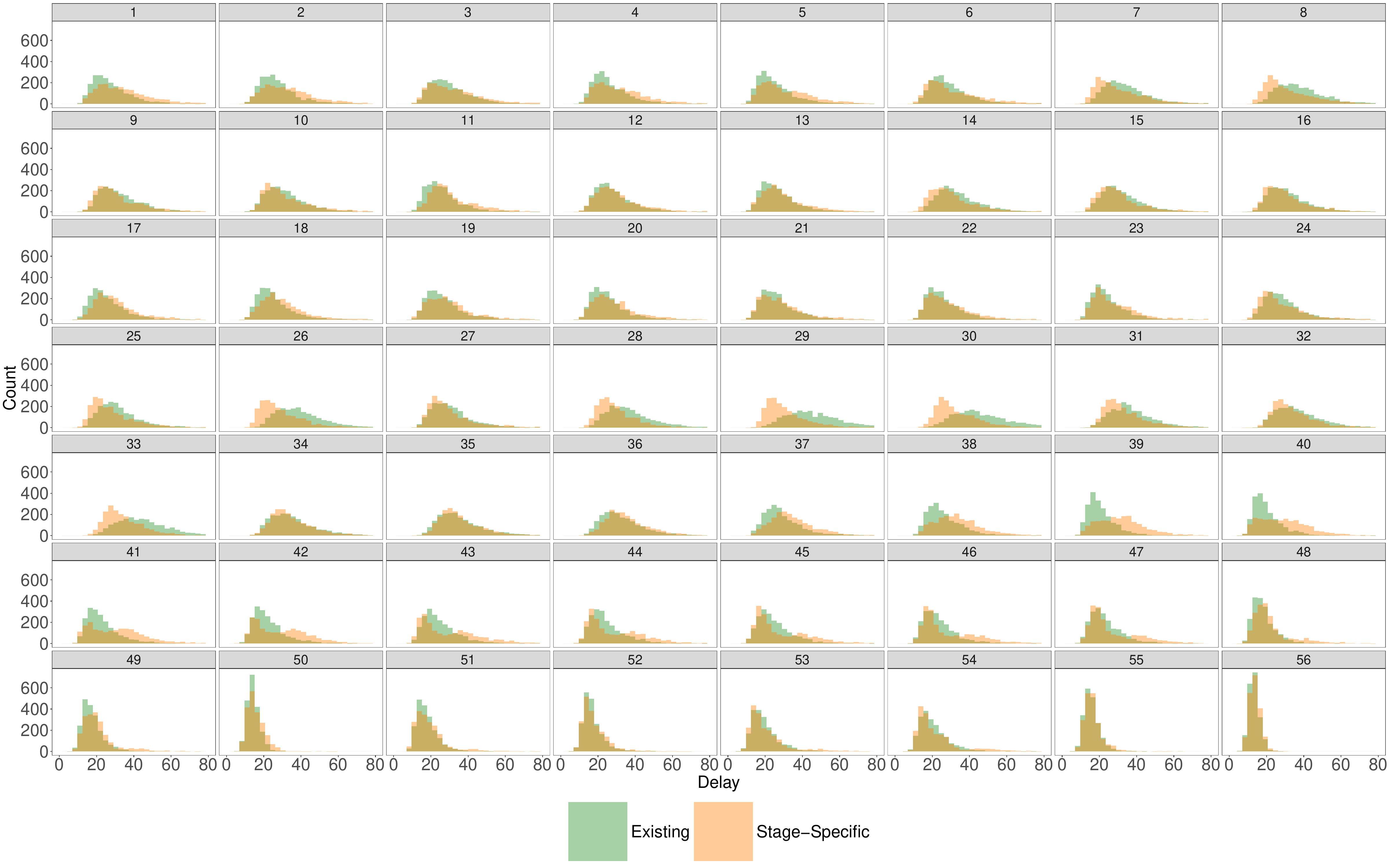}
\caption{\label{fig:sim2_both_gi_histogram} \footnotesize \textbf{Time-varying generation times in our simulated environment}. For individual calendar times, represented by numbers in facet labels, we present histograms of Monte Carlo samples from the generation time distribution under different temperature-dependent assumptions. For the existing (light blue) approach, we allow the transmission cycle to change every day based on the temperature observed that day. Our stage-specific approach means that the generation time distribution can vary based on temperatures observed for calendar times when the estimated (by Monte Carlo samples) generation events occurred.}
\end{figure}
\begin{figure}[t]
\centering
\includegraphics[scale = 0.17]{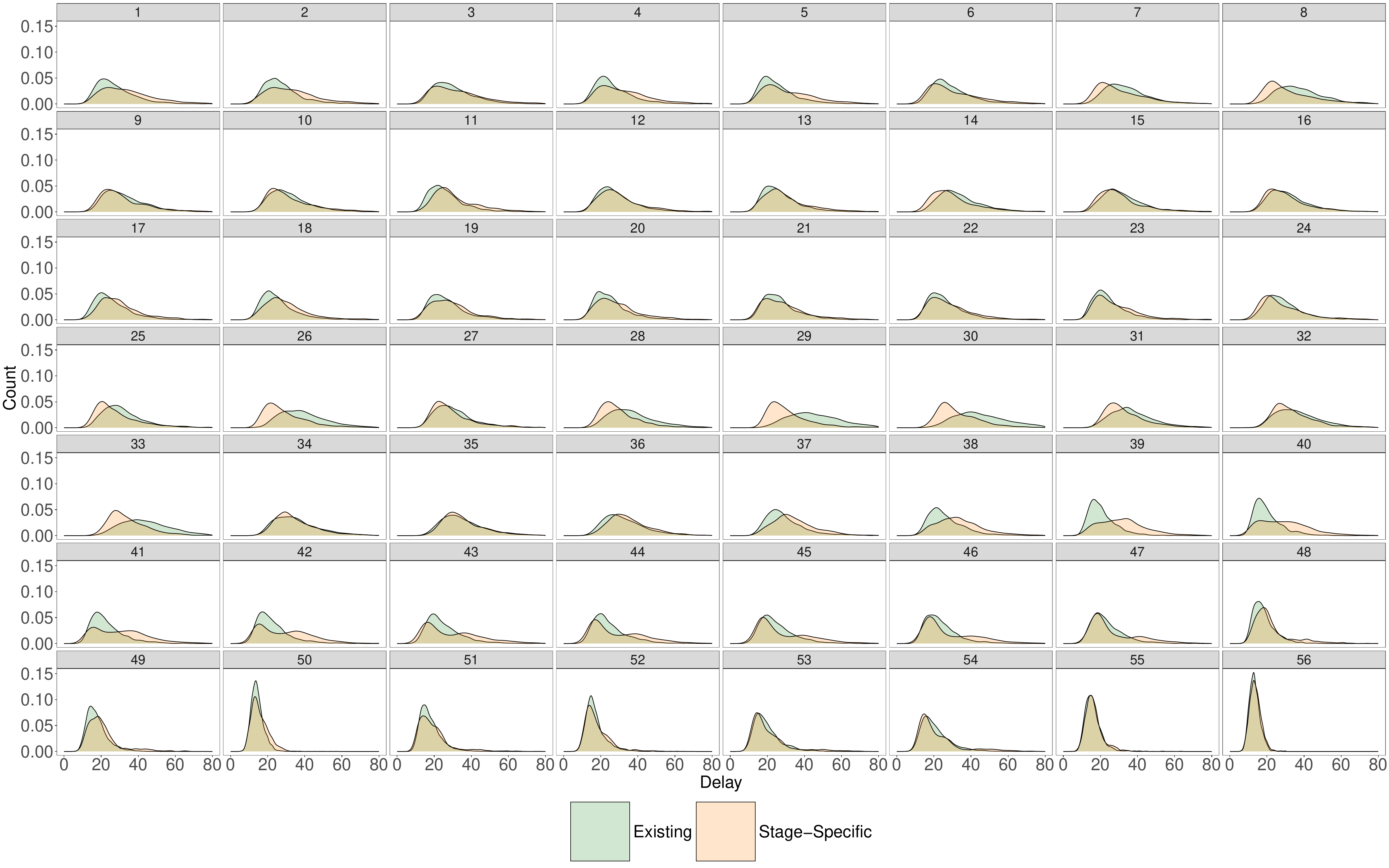}
\caption{\label{fig:sim2_both_gi_density} \footnotesize \textbf{Time-varying generation times in our simulated environment}. For individual calendar times, represented by numbers in facet labels, we present kernel density estimates of Monte Carlo samples from the generation time distribution under different temperature-dependent assumptions. For the existing (light blue) approach, we allow the transmission cycle to change every day based on the temperature observed that day. Our stage-specific approach means that the generation time distribution can vary based on temperatures observed for calendar times when the estimated (by Monte Carlo samples) generation events occurred.}
\end{figure}

In addition to Figure \ref{fig:application_gis_plot}, we present results for Ang Mo Kio, São João do Tauape, and the simulated environment in Figures \ref{fig:amk_both_gi_histogram}-\ref{fig:sim2_both_gi_density}.

\newpage

\newpage

\FloatBarrier
\newpage

\newpage

\newpage
\end{appendix}
\end{document}